\documentclass[10pt,a4paper]{article}
\usepackage[latin1]{inputenc}
\usepackage{ae,aecompl,amsbsy,amssymb,eurosym}
\usepackage{amsmath,amsthm}
\usepackage[dvips]{graphicx}
\usepackage{psfrag}
\usepackage{cases}
\usepackage{url}

 \numberwithin{equation}{section}
\newcommand\eq[1]{\begin{align}#1\end{align}}

\newcommand{\inte}{\!\int\!\!}
\newcommand{\ud}{\bar{d}}

\newcommand{\mm}{\mathcal M}

\newcommand{\mav}{\mathbf v}

\newcommand{\mar}{\mathbf r}
\newcommand{\maxx}{\mathbf x}

\newcommand{\mak}{\mathbf k}
\newcommand{\maB}{\mathbf B}
\newcommand{\maw}{\mathbf w}
\newcommand{\RR}{{\mathbb R}}
\newcommand{\CC}{{\mathbb C}}
\renewcommand{\wp}{P}

\title{\Large\bf{Dynamo effect in the Kraichnan magnetohydrodynamic turbulence}}
\author{Heikki Arponen\thanks{Helsinki University, Department of Mathematics and Statistics,
P.O. Box 68, 00014 Helsinki (Finland)} \ and Peter
Horvai\thanks{TNT group, Dipartimento di Fisica, Roma ``La
Sapienza'', P.zzle Aldo Moro, 2 I-00185 Roma (Italy)}}
\begin{document}
%
%\noindent
%
\maketitle

\begin{abstract}The existence of a dynamo effect in a simplified
magnetohydrodynamic model of turbulence is considered when the
magnetic Prandtl number approaches zero or infinity.  The magnetic
field is interacting with an incompressible Kraichnan-Kasantzev
model velocity field augmented with a viscous scale cutoff.  An
approximate system of equations in the different scaling ranges
can be formulated and solved, so that the solution tends to the
exact one when the viscous and magnetic-diffusive cutoffs approach
zero.  In this approximation we are able to determine analytically
the conditions for the existence of a dynamo effect and give an
estimate of the dynamo growth rate.  Among other things we show
that in the large Prandtl number case the dynamo effect is always
present.  Our analytical estimates are in good agreement with
previous numerical studies of the Kraichnan-Kasantzev dynamo by
Vincenzi~\cite{vincenzi}.

\end{abstract}

\section{Introduction}

As we are lacking in the complete understanding of turbulence, it
is useful to study simplified models which share many features of
the full problem. One class of models are the passive advection
models, where the velocity field is given some predetermined
statistics. One then studies the effect of this velocity field on
some other quantities such as the passive scalar, a temperature or
a dye density in the fluid (see e.g. \cite{kupiainen}). The term
passive refers to the absence of backreaction to the velocity
field. Naturally one would like to extend this study to passively
advected vector fields. There exist physically realistic models
such as the small resistivity magnetohydrodynamic equations for
the magnetic field $\maB \in \mathbb{R}^d$ interacting with a
fluid (see e.g. \cite{goldston, vincenzi, vergassola, paolo} and
references therein)

\eq{\partial_t \maB + \mav\cdot \nabla \maB - \maB \cdot \nabla
\mav = \kappa \Delta \maB .\label{equation}}
We will derive an equation for the pair correlation function

\eq{\big \langle B_i (t,\mar) B_j (t,\mar')\big \rangle}
averaged over the velocity statistics, and attempt to solve it
using a certain approximation scheme, which will be explained at
the end of this introduction. The $\mav$ above is the velocity
field of a conducting fluid. It is assumed to be incompressible as
in the constant density Navier-Stokes equation, that is, $\nabla
\cdot \mav =0$. Naturally the magnetic field is also
incompressible due to the absence of magnetic charges. $\kappa$ is
the resistivity divided by the vacuum permeability. The
``smallness'' of $\kappa$ here means that the equation is a good
approximation when the magnetic Reynolds number $R_M = V L
/\kappa$ is very large. Here $L$ is the integral scale of the
velocity field and $V$ is the r.m.s velocity at such a scale.
Since we assume from the beginning that $L \to \infty$, the
approximation holds. In reality we should also consider the
backreaction of $\maB$ on $\mav$, but we will only consider the
passive case and just assume $\mav$ to be given by the Kraichnan
statistics (see e.g. \cite{kupiainen} for a definition).  It is
defined as a Gaussian, mean zero velocity field with pair
correlation function
\eq{ \big\langle  v_i(t,\mar)v_j(t',\mar')\big\rangle =
\delta(t-t') D_0 \inte d \mak \frac{e^{i \mak \cdot (\mar -
\mar')}}{|\mak|^{d+\xi}} f(l_\nu|\mak|) P_{ij}(\mak)
 =: \delta(t-t') D_{ij}(\mar-\mar';l_\nu)\label{correlation}}
with $d \mak := \frac{d^d k}{(2 \pi)^3}$ and
\eq{P_{ij}(\mak)=\delta_{ij}-\frac{k_i k_j}{k^2}}
to guarantee incompressibility. It is evident that $D_{ij}$ is
homogenous and isotropic. The parameter $\xi$ for $0 \leq \xi \leq
2$  describes the roughness of the velocity field with $\xi=2/3$
corresponding to the Kolmogorov scaling. The function $f$ is an
ultraviolet cutoff, which simulates the effects of viscosity. It
decays faster than exponentially at large $k$, while $f(0)=1$ and
$f'(0)=0$. For example we could choose $f(l_\nu k)=\exp{(-l_\nu^2
k^2)}$, although the explicit form of the function is not needed
below. In the usual case without the cutoff function $f$ the velocity
correlation function behaves as a constant plus a term $\propto
r^\xi$, but in this case we have an additional scaling range for $ r
\ll l_\nu$ where it scales as $\propto r^2$. The length scale $l_\nu$
can be used to define a viscosity $\nu$ or alternatively one can use
$\kappa$ to define a length scale $l_\kappa$. We can then define the
Prandtl number\footnote{We choose to write the Prandtl number as $P$
instead of the usual $Pr$ since it appears so frequently in formulae.}
measuring the relative effects of viscosity and diffusivity as $\wp =
\nu/\kappa$. Note that the integral scale was assumed to be infinite,
i.e. there is no IR cutoff.\\

In addition to an advection term $\mav \cdot \nabla \maB$ in
Eq.~\eqref{equation}, familiar from the passive scalar problem,
there is also a stretching term $\maB \cdot \nabla \mav$. This,
and the fact that the magnetic field is a vector, will give rise
to some interesting deviations from the passive scalar case. The
most interesting one is probably the dynamo effect, an unbounded
growth of the magnetic field depending on the roughness of the
velocity field described by the parameter $\xi$, and the Prandtl
number.  This is in complete contrast to the passive scalar case,
where in the absence of external forcing the dynamics is always
dissipative \cite{hakulinen}.
\\

The dynamo effect has been previously studied by e.g. Kazantsev
\cite{kazantsev}, where he derived a Schrödinger equation for the
pair correlation function. However, the equation was still quite
difficult to analyze except in some special cases. Some analytical
and numerical results have been obtained e.g. in \cite{vergassola}
and \cite{vincenzi} (see the latter for further references). The
goal of the present paper is to extend these considerations by
introducing a set of approximate equations, which admit an exact
solution. The analysis proceeds along the same lines as in a
previous paper for a different problem by one of us \cite{peter}.
The problem in the analysis can be traced to existence of length
scales dividing the equation in different scaling ranges. In our
case there are two such length scales, one arising from the
diffusivity $\kappa$ and the other from the UV cutoff in the
velocity correlation function. As will be seen in appendix A, what
one actually needs in the analysis is the velocity
\emph{structure} function defined as

\eq{\frac{1}{2} \big\langle &
(v_i(t,\mar)-v_i(t,\mar'))(v_j(t',\mar)-v_j(t',\mar'))\big\rangle
\nonumber
\\ &= \delta(t-t') D_0 \inte \ud \mak \frac{1-e^{i \mak \cdot (\mar - \mar')}}{|\mak|^{d+\xi}} f(l_\nu|\mak|) P_{ij}(\mak)\nonumber
 \\ &=: \delta(t-t') d_{ij}(\mar-\mar';l_\nu)\label{structure}}
This is all one needs to derive a partial differential equation for
the pair correlation function of $\maB$, but it will still be very
difficult to analyze. Hence the approximation, which proceeds as
follows: \\

\textbf{1)} Consider the asymptotic cases where $r$ is far from
the length scales $l_\kappa$ and $l_\nu$ with the separation of
the length scales large as well. There are therefore three ranges
where the equation is simplified into a much more manageable form.
The equations are of the form $\partial_t H - \mathcal M H = 0$,
where $\mathcal M$ is a second order differential operator with
respect to the radial variable. We then consider the eigenvalue
problem $\mathcal M h = \lambda h$.\\

\textbf{2)} By a suitable choice of constant parameters in terms
of the length scales, we can adjust the differential equations to
match in different regions as closely as possible. Solving the
equations,
we obtain two independent solutions in all ranges.\\

\textbf{3)} We match the solutions by requiring continuity and
differentiability at the scales $l_\nu$ and $l_\kappa$. Also
appropriate boundary conditions
are applied.\\

\textbf{4)} According to standard physical lore, the form of
cutoffs do not affect the results when the cutoffs are removed. In
addition to $l_\nu$, we can interpret $l_\kappa$ as a cutoff.
Therefore we conjecture that the solution approaches the exact one
for small cutoffs. We also expect the qualitative results, such as
the existence of the dynamo effect, to apply for finite cutoffs as
well.\\

\begin{figure}
\begin{center}
\psfrag{a}{a)} \psfrag{b}{b)} \psfrag{c}{c)}
\psfrag{d}{d)}\psfrag{r}{r}
\includegraphics[width=0.8\textwidth]{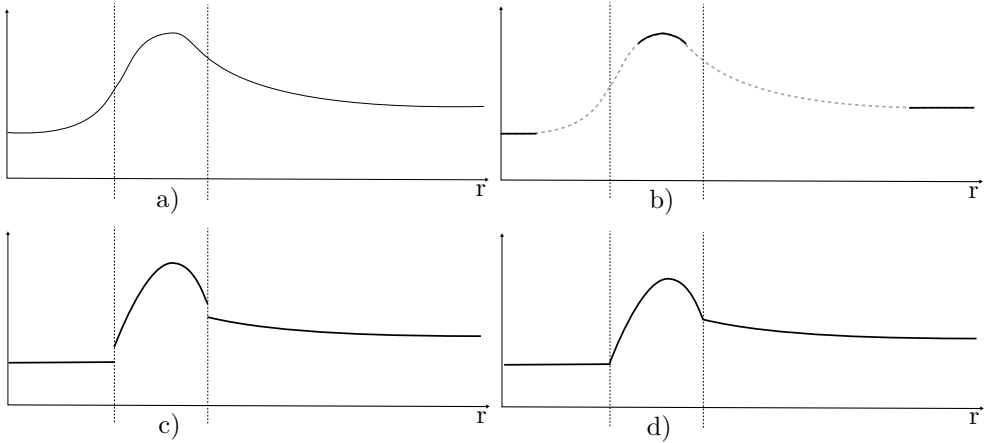}

\end{center}
\caption{A sketch of the procedure of approximating the example
equation. The dashed vertical lines correspond to either one of
the length scales $l_\nu$ and $l_\kappa$ with pictures a) a plot
of the ``real'' coefficient, which depends of the cutoff function
(and is really unknown), b) an approximate form obtained by taking
$r$ far from the length scales (dotted parts of the lines are
dropped) c) the approximations extended to cover all $r\in
\mathbb{R}$ and d) adjusting the coefficients to match at the
scales $l_\nu$ and $l_\kappa$. For $r$ much larger than the
cutoffs, the error due to the approximation is lost.}
\label{chopnjoin}
\end{figure}

For concreteness, suppose that $\mathcal M$ is of the form

\eq{\mathcal M = a(l_\nu,l_\kappa,r) \partial_r^2 +
b(l_\nu,l_\kappa,r) \partial_r +c(l_\nu,l_\kappa,r).}
The coefficients are some functions of the length scales $l_\nu$
and $l_\kappa$ and the radial variable $r$. In general, solving
the eigenvalue problem for such a differential equation is not
possible except numerically. However, we can approximate the
coefficients in the asymptotic regions when $r$ is far from the
length scales. The asymptotic coefficients are all power laws and
solving the equations becomes much easier.  Figure~\ref{chopnjoin}
illustrates this procedure corresponding to steps $1)$ and $2)
$ for any of the coefficients.\\

After some preparations, we begin by writing down the equation for
the pair correlation function of the magnetic field using the
It\^o formula. The derivation can be found in appendix A. The
equation is of third order in the radial variable, but it can be
manipulated into a second order equation by using the
incompressibility condition. In section 2 the approximate
equations will be derived when $\nu \ll \kappa$ and $\kappa \ll
\nu$, or Prandtl number small or large, respectively. We use
adimensional variables for sake of convenience and clarity. The
focus of the paper is mainly on the existence of the dynamo effect
and its growth rate. Therefore we consider the spectrum of
$\mathcal M$.  By a spectral mapping theorem, we relate the
spectra of $\mathcal M$ and the corresponding semigroup $e^{t
\mathcal M}$. It is then evident that if the spectrum of $\mathcal
M$ contains a positive part, there is exponential growth, i.e. a
dynamo effect.

\subsection{Structure function asymptotics}

Due to the viscous scale $l_\nu$ in the structure function
Eq.~\eqref{structure}, there are two extreme scaling ranges $r\gg
l_\nu$ (inertial range) and $r \ll l_\nu$. For $r \gg l_\nu$ we can
set $l_\nu = 0$ in Eq.~\eqref{structure} and obtain

\eq{ d_{ij}^>(\mar) := D_1 r^\xi \left( (d+\xi-1) \delta_{ij}-\xi
\frac{r_i r_j}{r^2}\right) ,\label{longstructure}}
where

\eq{D_1 = \frac{D_0 C_\infty}{(d-1)(d+2)}, \ \ \ \ \ C_\infty =
\frac{\Gamma(1-\xi/2)}{2^{d+\xi-2}\pi^{d/2}\Gamma (d/2 + \xi/2)}.}
The second case corresponds to the viscous range, which is to
leading order in $r$:

\eq{ d_{ij}^<(\mar):=  D_2 l_\nu^{\xi-2} r^2 \left(
(d+1)\delta_{ij} - 2 \frac{r_i r_j}{r^2} \right)
\label{shortstructure},}
where

\eq{D_2 =\frac{D_0 C_0}{(d-1)(d+2)}, \ \ \ \ \ C_0 = \inte \ud
\mak \frac{f(k)}{k^{d+\xi-2}}.}
We see that the viscous range form \eqref{shortstructure} can be
obtained from \eqref{longstructure} by a replacement $\xi \to 2$
and $D_1 \to D_2 l_\nu^{\xi-2}$.  Note that by adjusting the
cutoff
function $f$ we can also adjust $D_2/D_1$.\\

\subsection{Incompressibility condition}
Due to rotation and translation invariance, the equal-time
correlation function of $\maB$ must be of the form

\eq{ G_{ij}(t,|\maxx-\maxx'|):=\langle B_i(t,\maxx) B_j(t,\maxx')
\rangle = G_1(t,r) \delta_{i j} + G_2(t,r)\frac{r_i r_j}{r^2}
,\label{symmetry}}
where $r=|\maxx - \maxx'|$. Additional simplification arises from
the incompressibility condition $\partial_i G_{ij}=0$:

\eq{ \partial_r G_1 = - \frac{1}{r^{d-1}} \partial_r (r^{d-1} G_2)
. \label{incompressibility}}
The general solution of the incompressibility condition can be
written as

\eq{\left\{ \begin{array}{ll}
G_1 &= r H'(r) + (d-1) H(r) \\
G_2 &= -r H'(r)
\end{array} \right.\label{G1G2}}
In terms of a so far arbitrary function $H$. Alternatively, adding
the above equations we may write

\eq{H(r) = \frac{1}{d-1} \left( G_1 + G_2 \right).}
This observation leads to a considerable simplification in the
differential equation for the correlation function: whereas the
equations for $G_1$ and $G_2$ are of third order in $r$, we can
use the above result to obtain a second order equation for $H$.
Then we would get back to $G$ through Eqs.~\eqref{G1G2}; for example
we have for the trace of $G$:

\eq{G_{ii} = (d-1) \left( r H'(r) + d H(r) \right),}
although we refrain from doing this since $H$ has the same
spectral properties as $G_{ii}$.\\

\section{Equations of motion}

The equation of motion for the pair correlation function is
derived in appendix A:

\eq{\frac{d }{dt}G_{ij}(t,\mar) = 2 \kappa \Delta G_{ij}(t,\mar) +
d_{\alpha\beta} G_{ij,\alpha\beta}-d_{\alpha j,\beta} G_{i\beta,
\alpha} - d_{i \beta,\alpha} G_{\alpha j,\beta} +
d_{ij,\alpha\beta} G_{\alpha\beta}.}
The indices after commas are used to denote partial derivatives and we
use the Einstein summation. By taking $r \gg l_\nu$ and $r \ll l_\nu$
we can use the approximations \eqref{longstructure} and
\eqref{shortstructure} to write the equation in the corresponding
ranges. This is done for the quantity $H=(G_1 + G_2)/(d-1)$ in the
Appendix~\ref{apx:PDE-2-B} as well, resulting in the equations

\eq{\frac{d H}{dt} &= \xi (d-1)(d+\xi) D_1 r^{\xi-2} H + \left[
2(d+1)\kappa +(d^2 -1+2\xi)D_1 r^\xi \right]  \frac{1}{r} H'
\nonumber
\\ &+\left[ 2 \kappa + (d-1) D_1 r^\xi  \right] H'', \ \ \ r \gg
l_\nu.\label{longeq}}
and

\eq{\frac{d H}{dt} &= 2 (d-1)(d+2) D_2 l_\nu^{\xi-2} H + \left[
2(d+1)\kappa +(d^2 +3)D_2 l_\nu^{\xi-2} r^2  \right] \frac{1}{r}
H' \nonumber
\\ &+\left[ 2 \kappa + (d-1) D_2 l_\nu^{\xi-2} r^2  \right] H'', \ \ \
r \ll l_\nu.\label{shorteq}}
Simple dimensional analysis leads to the observation

\eq{[\kappa] = [D_1 r^\xi] = [D_2 l_\nu^{\xi-2} r^2],}
where the brackets denote the scaling dimension of the quantities.
We define the length scale $l_\kappa$ as the scale below which the
diffusive effects of $\kappa$ become important. This will be done
explicitly below for different Prandtl number cases. In general,
one can write $\kappa = D_1 l_\nu^{\xi -p} l_\kappa^p$ for some $p
\in (0,2]$. Now one just needs to identify the dominant terms in
the three scales divided by $l_\nu$ and $l_\kappa$. For sake of
clarity, we choose to write these equations in adimensional
variables. This can be done for example by defining $r= l \rho$
and $t=l^{2-\xi} \tau /D_1$ with $l$ being a length scale. It
turns out to be convenient to choose the larger of $l_\kappa$ and
$l_\nu$ as $l$. Since we deal with a stochastic velocity field
with no
 intrinsic dynamics, we cannot, in principle, talk about viscosity.
   However, it is convenient to define a viscosity $\nu$ (of dimension
    length squared divided by time) by dimensional analysis from the
     length scale $l_\nu$ and the dimensional velocity magnitude $D_1$,
      giving a relationship between $\nu$, $l_\nu$ and $D_1$ similar to
       what we would get in a dynamical model. We therefore define

\eq{\nu := D_1 l_\nu^\xi.}
 This
        permits us to define the Prandtl number in the standard manner
         as $P = \nu/\kappa$. We then consider the
cases $\wp \ll 1$ and $\wp \gg 1$.

\subsection{Small Prandtl number}
\begin{figure}[h]
\begin{center}
\psfrag{a}{$l_\nu$} \psfrag{b}{$l_\kappa$} \psfrag{c}{$l_\nu /
l_\kappa$} \psfrag{d}{1}\psfrag{r}{$r$}
\psfrag{o}{$\rho$}\psfrag{S}{S}\psfrag{M}{M}\psfrag{L}{L}
\includegraphics[width=0.5\textwidth]{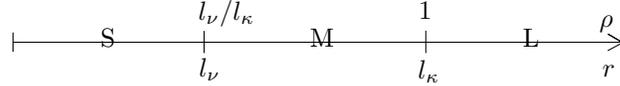}

\end{center}
\caption{Sketch of the scaling ranges at small Prandtl number.}
\label{smallpr}
\end{figure}
Now $\nu \ll \kappa$, and we choose as adimensional variables

\eq{\left\{ \begin{array}{ll}
r&=l_\kappa \rho  \\
t&=\frac{l_\kappa^{2-\xi}}{D_1} \tau.
\end{array} \right. \label{eq:adim-small-P}}
Note that the relation between $l_\kappa$ and $\kappa$ has not yet
been determined.  In these variables the equations \eqref{longeq} and
\eqref{shorteq} become

\eq{\partial_\tau H &= \xi (d-1)(d+\xi) \rho^{-2+\xi} H +
\left[2(d+1) \frac{\kappa}{D_1 l_\kappa^\xi} + (d^2 -1 + 2 \xi)
\rho^\xi \right]\frac{1}{\rho} \partial_\rho H  \nonumber \\
&+ \left[ 2 \frac{\kappa}{D_1 l_\kappa^\xi} +
(d-1)\rho^\xi\right]\partial_\rho^2 H , \ \ \rho \gg l_\nu /
l_\kappa \label{eka}}
and

\eq{\partial_\tau H &= 2 (d-1)(d+2) \frac{D_2}{D_1} \left(
\frac{l_\nu}{l_\kappa} \right)^{\xi-2} H + \left[2(d+1)
\frac{\kappa}{D_1 l_\kappa^\xi} + (d^2 +3) \frac{D_2}{D_1} \left(
\frac{l_\nu}{l_\kappa} \right)^{\xi-2}
\rho^2 \right]\frac{1}{\rho} \partial_\rho H  \nonumber \\
&+ \left[ 2 \frac{\kappa}{D_1 l_\kappa^\xi} + (d-1)
\frac{D_2}{D_1} \left( \frac{l_\nu}{l_\kappa} \right)^{\xi-2}
\rho^2\right]\partial_\rho^2 H ,\ \ \rho \ll l_\nu /
l_\kappa.\label{toka}}
As mentioned above, we also consider $r \ll l_\kappa$ and $r \gg
l_\kappa$, that is $\rho \ll 1$ and $\rho \gg 1$, respectively.  There
are now three regions in $\rho$, divided by $l_\nu / l_\kappa$ and
$1$, with $l_\nu /l_\kappa \ll 1$. The regions, solutions and various
other quantities will be labelled by $S$, $M$ and $L$, corresponding
to $\rho \ll l_\nu / l_\kappa$, $l_\nu / l_\kappa \ll \rho \ll 1$ and
$1 \ll \rho$. See Fig.~\ref{smallpr} for quick reference.  Therefore
the short range equation will be derived from Eq.~\eqref{toka} and the
two others from \eqref{eka}.  Consider for example explicitly the
coefficients of $\partial_\rho^2 H$:

\eq{L: \ \ \  &2 \frac{\kappa}{D_1 l_\kappa^\xi} + (d-1)\rho^\xi
\nonumber \\ M: \ \ \ & 2 \frac{\kappa}{D_1 l_\kappa^\xi} +
(d-1)\rho^\xi \\ S: \ \ \ &2 \frac{\kappa}{D_1 l_\kappa^\xi} +
(d-1) \frac{D_2}{D_1} \left( \frac{l_\nu}{l_\kappa}
\right)^{\xi-2} \rho^2. \nonumber}
By definition of the length scale $l_\kappa$, in the region $L$
the diffusivity is negligible and in the region $M$ it is
dominant, as it is in the region $S$ since in there $\rho$
approaches zero. The coefficients are then approximately

\eq{L: \ \ \  &(d-1)\rho^\xi \nonumber \\ M: \ \ \ & 2
\frac{\kappa}{D_1 l_\kappa^\xi} \\ S: \ \ \ &2 \frac{\kappa}{D_1
l_\kappa^\xi}. \nonumber}
Matching the coefficients of $L$, $M$ at $\rho=1$ provides us with
a condition (matching between $S$ and $M$ gives nothing new)

\eq{d-1 = 2 \frac{\kappa}{D_1 l_\kappa^\xi}.\label{eq:P0-match-ML}}
This is used as a definition of $\kappa$ as $\kappa = \frac{1}{2}
(d-1) D_1 l_\kappa^\xi$. Writing down the short range equation
with the above approximations,

\eq{\partial_\tau H_S = 2(d-1)(d+2) \frac{D_2}{D_1}\left(
\frac{l_\nu}{l_\kappa} \right)^{\xi-2} H_S + (d^2 -1)
\frac{1}{\rho} \partial_\rho H_S + (d-1)\partial_\rho^2 H_S,}
by using the derived expression for the Prandtl number,

\eq{\wp = \frac{\nu}{\kappa} = \frac{2}{d-1} \left(
\frac{l_\nu}{l_\kappa} \right)^\xi,\label{smallprnumber}}
and by defining

\eq{\frac{D_2}{D_1} =  \left( \frac{2}{d-1}\right)^{1-2/\xi}}
(remember that $D_2$ could be adjusted by a choice of the cutoff
function $f$, see Eq.~\eqref{shortstructure} and below) a more
neat expression is obtained for the short range equation. We can
now write down all the equations:
\begin{subnumcases}{}
  \label{SmallPrEquationsa}\partial_\tau H_S
&\hspace{-.6cm}$=\
 2 (d-1)(d+2)
\wp^{1-2/\xi} H_S + (d^2-1) \frac{1}{\rho}
\partial_\rho H_S  + (d-1) \partial_\rho^2 H_S
$
\\[3pt]
\label{SmallPrEquationsb}
  \partial_\tau H_M
&\hspace{-.6cm}$=\
  \xi (d-1)(d+\xi) \rho^{-2+\xi} H_M +
(d^2-1)  \frac{1}{\rho}
\partial_\rho H_M+ (d-1) \partial_\rho^2 H_M
$
\\[3pt]
\label{SmallPrEquationsc}
  \partial_\tau H_L
&\hspace{-.6cm}$=\
  \xi (d-1)(d+\xi) \rho^{-2+\xi} H_L +
 (d^2 -1 + 2 \xi)
\rho^{\xi-1} \partial_\rho H_L + (d-1)\rho^\xi\partial_\rho^2 H_L
 $ .
\end{subnumcases}
\subsection{Large Prandtl number}

\begin{figure}[h]
\begin{center}
\psfrag{a}{$l_\kappa$} \psfrag{b}{$l_\nu$} \psfrag{c}{$l_\kappa /
l_\nu$} \psfrag{d}{1}\psfrag{r}{$r$}
\psfrag{o}{$\rho$}\psfrag{S}{S}\psfrag{M}{M}\psfrag{L}{L}
\includegraphics[width=0.5\textwidth]{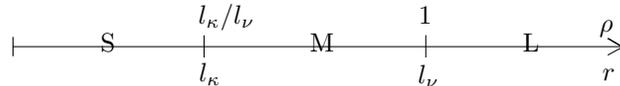}

\end{center}
\caption{Sketch of the scaling ranges at large Prandtl number.}
\label{largepr}
\end{figure}
Now $\nu \gg \kappa$, and we choose

\eq{\left\{ \begin{array}{ll}
r&=l_\nu \rho  \\
t&=\frac{l_\nu^{2-\xi}}{D_1} \tau.
\end{array} \right.\label{eq:adim-large-P}}
Then the equations \eqref{longeq} and \eqref{shorteq} for $r \gg
l_\nu$ and $r \ll l_\nu$ become in the new variables

\eq{\partial_\tau H &= \xi (d-1)(d+\xi) \rho^{-2+\xi} H +
\left[2(d+1) \frac{\kappa}{D_1 l_\nu^\xi} + (d^2 -1 + 2 \xi)
\rho^\xi \right]\frac{1}{\rho} \partial_\rho H  \nonumber \\
&+ \left[ 2 \frac{\kappa}{D_1 l_\nu^\xi} +
(d-1)\rho^\xi\right]\partial_\rho^2 H , \ \ \rho \gg 1
\label{eka2}}
and

\eq{\partial_\tau H &= 2 (d-1)(d+2) \frac{D_2}{D_1} H +
\left[2(d+1) \frac{\kappa}{D_1 l_\nu^\xi} + (d^2 +3)
\frac{D_2}{D_1}
\rho^2 \right]\frac{1}{\rho} \partial_\rho H  \nonumber \\
&+ \left[ 2 \frac{\kappa}{D_1 l_\nu^\xi} + (d-1) \frac{D_2}{D_1}
\rho^2\right]\partial_\rho^2 H ,\ \ \rho \ll 1.\label{toka2}}
The ranges $S$, $M$ and $L$ now correspond to $\rho \ll l_\kappa /
l_\nu $, $l_\kappa / l_\nu \ll \rho \ll 1$ and $ 1 \ll \rho$, see
Fig.~\ref{largepr}. Note that equations in both $S$ and $M$ are now
derived from Eq.~\eqref{toka2}.  As before, we consider again the
coefficients of $\partial_\rho^2 H$ and drop the terms $\propto
\kappa$ in $L$ and $\propto \rho^2$ in $S$. The diffusive effects are
not dominant in the region $M$ since $r \gg l_\kappa$, so we drop the
$\propto \kappa$ term in $M$ too. The approximative coefficients are
then

\eq{L: \ \ \  &  (d-1)\rho^\xi \nonumber \\ M: \ \ \ & (d-1)
\frac{D_2}{D_1} \rho^2 \\ S: \ \ \ &2 \frac{\kappa}{D_1
l_\nu^\xi}.}
We then obtain two equations by matching the coefficient of $L$
with $M$ at $\rho = 1$ and of $M$ with $S$ at $l_\kappa / l_\nu$:

\eq{\frac{D_2}{D_1}(d-1)&= (d-1) ,\nonumber \\
(d-1) \frac{D_2}{D_1} \left( \frac{l_\kappa}{l_\nu}\right)^2 &= 2
\frac{\kappa}{D_1 l_\nu^\xi} }
with solutions

\eq{D_2 &= D_1 \nonumber \\
\kappa &= \frac{d-1}{2} D_1 l_\kappa^2 l_\nu^{\xi-2}.}
The Prandtl number is in this case

\eq{\wp = \frac{2}{d-1} \left( \frac{l_\nu}{l_\kappa}
\right)^2.\label{largeprnumber}}
Note that one can obtain this from the small Prandtl number
equation \eqref{smallprnumber} by replacing $\xi \to 2$. This is a
reflection of a more subtle observation that the large Prandtl
number case for \emph{any} $\xi$ is similar to the small Prandtl
number case with $\xi = 2$. We collect the equations using the
above approximations,
\begin{subnumcases}{}
\label{prgg1eqsa}
  \partial_\tau H_S
&\hspace{-.6cm}$=\
  2 (d-1)(d+2)  H_S +
  2 \frac{d+1}{\wp }\frac{1}{\rho}\partial_\rho H_S +
  \frac{2}{\wp} \partial_\rho^2 H_S
$
\\[3pt]
  \partial_\tau H_M
&\hspace{-.6cm}$=\
  2 (d-1)(d+2)  H_M +
  (d^2 +3)\rho\partial_\rho H_M +
  (d-1) \rho^2 \partial_\rho^2 H_M \label{prgg1eqsb}
$
\\[3pt]
\label{prgg1eqsc}
  \partial_\tau H_L
&\hspace{-.6cm}$=\
  \xi (d-1)(d+\xi) \rho^{-2+\xi} H_L +
  (d^2 -1 + 2 \xi)\rho^{\xi-1} \partial_\rho H_L +
  (d-1)\rho^\xi\partial_\rho^2 H_L
$ .
\end{subnumcases}
Note that the short and long range equations are somewhat similar
to the respective small Prandtl number ones,
Eqs.~(\ref{SmallPrEquationsa},\ref{SmallPrEquationsb},\ref{SmallPrEquationsc}).
However, the equation in the medium range above is scale invariant
in $\rho$, unlike the corresponding small Prandtl number one.

\section{Resolvent}

Given a differential operator $\mathcal M$ with a domain $D(\mm)$,
we define the resolvent

\eq{R(z,\mm):= (z-\mm)^{-1}}
and the resolvent set as

\eq{\rho(\mm):= \left\{ z \in \mathbb{C}| z-\mm : D(\mm) \to
\textnormal{X is bijective} \right\}.}
The complement of the resolvent set, denoted by $\sigma(\mm)$, is the
spectrum of $\mm$. According to the Hille-Yosida theorems (see
e.g. \cite{engel}), if $(\mm, D(\mm))$ is closed and densely defined
and if there exists $z_0 \in \mathbb{R}$ such that for each $z \in
\mathbb{C}$ with $\Re{z} > z_0$ we have $z \in \rho(\mm)$, and

\eq{\parallel R(z,\mm)\parallel \leq \frac{1}{\Re{z}-z_0},}
then $\mm$ is the generator of a strongly continuous semigroup
$T(t)$ satisfying

\eq{\parallel T(t) \parallel \leq e^{z_0 t}\label{semigroupbound}}
and vice versa.  If in addition $\mm$ is a so called sectorial
operator, meaning that its spectrum is contained in some angular
sector $\{ z\in\mathbb{C} : |\arg (z-z_0)| > \alpha > \pi/2\}$ and
that outside this sector the resolvent satisfies the (stronger)
estimate

\eq{\parallel R(z,\mm)\parallel \leq \frac{C}{|z-z_0|},
\label{ineq:anly-resolv}}
then $\mm$ generates an analytical semigroup, for which the spectral
mapping theorem

\eq{\sigma \left( T(t)\right) = \{ 0 \} \cup e^{t \sigma (\mm)}.}
holds, relating the spectrum of the generator to that of the
semigroup.  One can also use the Cauchy integral formula

\eq{T(t) := e^{t \mm} = \frac{1}{2 \pi i} \int\limits_\mathcal{C}
dz e^{z t} R(z,\mm),}
where the contour surrounds the spectrum $\sigma (\mm)$.

We do not prove here that $\mm$ is sectorial, however we refer the
interested reader to the general mathematical theory in \cite{Lunardi}
where it is explained and substantiated that strongly elliptic
operators are, under quite general assumption, sectorial generators,
on a wide range of Banach spaces (e.g.\ $L^p$ and $C^1$ spaces to name
but a few).

According to the above discussion, in order to explain the existence
of the dynamo effect and its growth rate, we only need to find the
spectrum of $\mm$ via the resolvent set $\rho (\mm)$.  Note that we
are interested only in the positive part of the spectrum, since we
want to determine the existence of the dynamo effect only.

%% We will simply assume that the semigroup satisfies the bound
%% (\ref{semigroupbound}) because of physical reasons (finite growth
%% rate).  The semigroup can also be seen to be analytic such that also
%% the spectral mapping theorem holds.

The operator $\mm$ in our case is cut up as the operators $\mm_L$,
$\mm_M$ and $\mm_S$ in the corresponding ranges, obtained from the
equations (\ref{SmallPrEquationsa}, \ref{SmallPrEquationsb},
\ref{SmallPrEquationsc}) and (\ref{prgg1eqsa}, \ref{prgg1eqsb},
\ref{prgg1eqsc}).  The resolvent is found from the equation

\eq{\left(z-\mm \right) R(z,\mm) (\rho, \rho') = \delta (\rho,
\rho').}
Since we are primarily interested in the long range ($L$) behavior
$\rho>1$, we let $\rho'$ stay in the region $L$ at all times.  This
results in three equations

\eq{\left\{ \begin{array}{ll} \left( z-\mathcal M_L \right) R_L
(\rho,\rho') &= \delta (\rho-\rho')
\\
\left( z-\mathcal M_M \right) R_M (\rho,\rho') &= 0
\\
\left( z-\mathcal M_S \right) R_S (\rho,\rho') &= 0
\end{array} \right. \label{resolvents}}
where $R_L(\rho,\rho')$ is the expression of the resolvent for $\rho
\in L$ (the large scale range) and $\rho' \in \mathbb{R}_+$ and
similarly $R_M$ and $R_S$ are valid when $\rho$ is in the middle and
small scale ranges respectively.  We require the following boundary
conditions from the resolvents: for small $\rho$ we are in the
diffusion dominated range, so we require smooth behavior at $\rho \to
0$. For large $\rho$ we eventually cross the integral scale (although
we haven't defined it explicitly) above which the velocity field
behaves like the $\xi=0$ Kraichnan model, leading to diffusive
behavior at the largest scales for which the appropriate condition on
the resolvent is exponential decay at infinity.

\subsection{Solutions}

Assuming $\rho \neq \rho'$, we solve the equations \eqref{resolvents}
with the corrsponding operators $\mathcal M$.

The operator ${\mathcal M}_L$ does not depend on the Prandtl number.
So in the region $L$, we get from e.g. Eq.~\eqref{prgg1eqsc} (we use
lowercase letters $h^\pm$ to denote the independent solutions)
\eq{h_L^\pm (\rho)=\rho^{-d/2 -\frac{\xi}{d-1}} \tilde Z_{\lambda}^\pm
\left( w \rho^{1-\xi/2} \right),\label{hLpm}}
where $\tilde Z_\lambda ^+ \equiv I_\lambda$ and $\tilde Z_\lambda^-
\equiv K_\lambda$ are modified Bessel functions of the first and
second kind respectively, and we have introduced
\eq{w = \frac{2}{2-\xi}\sqrt{\frac{z}{(d-1)}}.\label{w}}
and the order parameter $\lambda$ is

\eq{
  \lambda
=
  \frac{\sqrt{d [2(d-1)^3 - (d-2)(2\xi+d-1)^2]}}
       {(2-\xi)(d-1)}
.
\label{lambda}
}

%% \eq{\lambda = \frac{\sqrt d}{2-\xi}\sqrt{d- 4
%% (d-2)\frac{\xi}{d-1}\left(
%% 1+\frac{\xi}{d-1}\right)}\label{lambda}.}
%
%

Because the range $S$ is always in the diffusive region, we require
smoothness of the solution at zero. Only one of the solutions
satisfies this, so we get from Eqs.~\eqref{SmallPrEquationsa} and
\eqref{prgg1eqsa}

\eq{h_{S,1} (\rho) = \rho^{-d/2} \ I_{d/2} \left( \sqrt{
\frac{z}{d-1}-2 (d+2) \wp^{1-2/\xi}} \rho \right)}
and \eq{h_{S,2} (\rho) = \rho^{-d/2} \ I_{d/2} \left( \sqrt{
\frac{z}{2}-(d-1)(d+2)} \sqrt \wp \ \rho \right),}
where the subindex $1$ refers to $\wp \ll 1$ (small Prandtl number)
and $2$ to $\wp \gg 1$. We will use this notation in other objects as
well.\\

In the range M, when $\wp \gg 1$ we have the scale invariant equation
in \eqref{prgg1eqsb} with power law solutions

\eq{h_{M,2}^\pm (\rho)= \rho^{-d/2 -\frac{2}{d-1} \pm \zeta},}
where
\begin{equation}
\label{def:zeta}
  \zeta
=
  \sqrt{\frac{z-z_2}{d-1}}
\end{equation}
with
\begin{equation}
\label{def:z_2}
  z_2
=
  -\frac{d-1}{4}[(2-\xi)\lambda]^2|_{\xi=2}
=
 -\frac{d}{4(d-1)} (d^3-10d^2+9d+16)
\,.
\end{equation}

%% \eq{\zeta = \frac{\sqrt{d(d^3-10d^2+9d+16) +4(d-1) z}}{2(d-1)}.}

The medium range equation for $\wp \ll 1$ cannot be solved exactly,
but we can consider it in two different asymptotic cases.  From
Eq.~\eqref{SmallPrEquationsb} we get

\eq{  \left(\xi (d+\xi) \rho^{-2+\xi}-\frac{z}{d-1}\right) R_M +
(d+1) \frac{1}{\rho}
\partial_\rho R_M+  \partial_\rho^2 R_M = 0\label{3.13}}
and note that since by definition of the medium range $l_\nu /
l_\kappa \ll \rho \ll 1 $, i.e.\ $1 <
\rho^{-2+\xi} < (l_\kappa / l_\nu)^{2-\xi}$ (the $\ll$ was replaced by
$<$, so that things remain valid even as $\xi\to2$), the term $\propto
\rho^{-2 + \xi}$ can be dropped if we assume that
\begin{equation}
\label{ineq:z-P-large}
  |z|
>
  (l_\kappa / l_\nu)^{2-\xi}
\approx
  P^{-\frac{2-\xi}{\xi}}
\,.
\end{equation}
If on the other hand we have
\begin{equation}
\label{ineq:z-P-small}
  |z|<  1
\,,
\end{equation}
then $z$ can be neglected in the equation.

The solution for large $z$ is similar to the short range solutions,

\eq{h_{M,1}^{\pm} (\rho) = \rho^{-d/2} \tilde Z^{\pm}_{d/2} \left(
\sqrt{\frac{z}{d-1}} \rho \right), \ \ \ \ |z| > (l_\kappa /
l_\nu)^{2-\xi}, \label{largez}}
where we denoted the $\wp \ll 1$ case by a subscript $1$.  For small
$z$ we have instead

\eq{h_{M,1}^\pm = \rho^{-d/2} Z^{\pm}_{d/\xi} \left( 2 \sqrt{d/\xi
+1} \rho^{\xi/2} \right), \ \ \ \ |z| < 1.\label{smallz}}
where $Z_{d/\xi}^+ \equiv J_{d/\xi}$ and $Z_{d/\xi}^- \equiv
Y_{d/\xi}$ are Bessel functions of the first and second kind
respectively.  It turns out however that the explicit form of the
above solutions affects only a specific numerical multiplier and has
no effect on the presence of the dynamo. Because of this we in fact
derive a lower bound for the growth rate which in view of the present
approximation provides a more reliable result.

\subsection{Matching of solutions}

Consider equations \eqref{resolvents}.  We denote the long range
regions $\rho<\rho'$ and $\rho>\rho'$ as $L_<$ and $L_>$. The boundary
conditions for the resolvent demanded finiteness at $\rho =0$, but in
general the resolvent must be in $L_2(\mathbb{R}_+)$.  We therefore
have in the region $L_>$ only the $h_L^-$ solution, since it decays as
a stretched exponential at infinity (the other one grows as a
stretched exponential). We also drop the subscripts labelling the
different Prandtl number cases for now. The full solutions are written
as follows:

\eq{\left\{ \begin{array}{ll} R_{S} (z|\rho,\rho') &=\alpha
 h_{S} (\rho)
\\
R_{M} (z|\rho,\rho') &=C_{M}^+ h_{M}^+ (\rho) + C_{M}^- h_{M}^-
(\rho)
\\
R_{L_<} (z|\rho,\rho') &= C_L^+ h_L^+(\rho)+C_L^- h_L^-(\rho)
\\
R_{L_>}(z|\rho,\rho') &= \beta h_L^- (\rho),
\end{array}\right.
}
We denote the matching point between the short and medium ranges
by $a_i$, i.e.

\eq{\left\{\begin{array}{ll} a_1 &= l_\nu/l_\kappa   \\ a_2 &=
l_\kappa / l_\nu  \end{array} \right.\label{aANDb}}
The other matching points are $\rho=1$ and $\rho=\rho'$ in both
cases. There are six coefficients to be determined, $\alpha,
C_M^\pm, C_L^\pm$ and $\beta$, and in total six conditions, four
from the continuity and differentiability at $\rho= a_i$ and
$\rho=1$ and two conditions at $\rho=\rho'$ around the delta
function, so all coefficients will be determined from these. They
will then depend on the variables $z$ and $\rho'$. The $\mathrm{C}^1$
conditions at $\rho = 1$ are

\eq{C_L^+ h_L^+ (1) + C_L^- h_L^- (1) = C_{M}^+ h_{M}^+
(1)+C_{M}^- h_{M}^- (1)}
and

\eq{C_L^+ \partial h_L^+ (1) + C_L^- \partial h_L^- (1) = C_{M}^+
\partial h_{M}^+ (1)+C_{M}^- \partial h_{M}^- (1),}
where we denoted $\partial h (1) = \partial_\rho h(\rho)
|_{\rho=1}$. This can be expressed conveniently as

\eq{\begin{pmatrix}
h_L^+ &\!\!\! h_L^- \\
\partial h_L^+ &\!\!\! \partial h_L^-
\end{pmatrix}_{1} \begin{pmatrix} C_L^+ \\ C_L^- \end{pmatrix} = \begin{pmatrix}
h_{M}^+ &\!\!\! h_{M}^- \\
\partial h_{M}^+ &\!\!\! \partial h_{M}^-
\end{pmatrix}_{1} \begin{pmatrix} C_{M}^+ \\ C_{M}^- \end{pmatrix}}
where the matrix subindex refers to evaluation of the matrix elements
at $\rho = 1$.  Since we have only one solution at short range, we get
similarly at $a_i$

\eq{\begin{pmatrix}
h_{M}^+ &\!\!\! h_{M}^- \\
\partial h_{M}^+ &\!\!\! \partial h_{M}^-
\end{pmatrix}_{a_i} \begin{pmatrix} C_{M}^+
\\ C_{M}^- \end{pmatrix} = \alpha \begin{pmatrix} h_{S}
\\ \partial h_{S} \end{pmatrix}_{a_i} .}
where again the matrix subindex indicates the point where matrix
elements are to be evaluated.  We can solve these for $C_L^\pm$,

\eq{\begin{pmatrix} C_L^+
\\ C_L^- \end{pmatrix} = \mathcal J' \begin{pmatrix}
\partial h_L^- &\!\!\! -h_L^- \\
-\partial h_L^+ &\!\!\!h_L^+
\end{pmatrix}_{1} \begin{pmatrix}
h_{M}^+ &\!\!\! h_{M}^- \\
\partial h_{M}^+ &\!\!\! \partial h_{M}^-
\end{pmatrix}_{1} \begin{pmatrix}
\partial h_{M}^- &\!\!\! -h_{M}^- \\
-\partial h_{M}^+ &\!\!\! h_{M}^+
\end{pmatrix}_{a_i} \begin{pmatrix} h_{S}
\\ \partial h_{S} \end{pmatrix}_{a_i}.\label{cLmatrix}}
The numeric constant $\mathcal J'$ above contains the determinants of
the inverted matrices and $\alpha$. It is certainly nonsingular due to
the linear independence of the solutions. We have decided not to
explicitly write it down since, as we will see below, we only need the
fraction $C_L^- / C_L^+$. Now we have piecewise the resolvents

\eq{\left\{ \begin{array}{ll} R_{L_<} (z|\rho,\rho') &= C_L^+
\left(h_L^+(\rho)+\frac{C_L^-}{C_L^+} h_L^-(\rho)\right)
\\
R_{L_>}(z|\rho,\rho') &= \beta h_L^- (\rho),
\end{array}\right.\label{GL1andGL2}
}
and we still need to use the first equation of \eqref{resolvents} for
$C_L^+$ and $\beta$. The continuity condition is

\eq{C_L^+ \left(
h_L^+(z,\rho')+\frac{C_L^-}{C_L^+}h_L^-(z,\rho')\right) = \beta
h_L^- (z,\rho').}
The other condition is obtained by integrating the equation with
respect to $\rho$ over a small interval and then shrinking the
interval to zero:

\eq{C_L^+ \left(
\partial h_L^+(\rho')+\frac{C_L^-}{C_L^+}\partial h_L^-(\rho')\right)
- \beta \partial h_L^-(\rho')=1.}
These can be solved to yield

\eq{C_L^+ = \frac{h_L^- (\rho')}{\mathcal W (h_L^+,h_L^-)(\rho')
}}
and
\eq{\beta = \frac{C_L^+ h_L^+ (\rho') + C_L^- h_L^- (\rho')
}{C_L^+ \mathcal W (h_L^+,h_L^-)(\rho') },}
where $\mathcal W$ is the Wronskian, $\mathcal W (f,g)= f g' - f' g$.
Explicitly from Eq.~\eqref{hLpm},

\eq{\mathcal W (h_L^+,h_L^-)(z,\rho') =
(\rho')^{-d-\frac{2\xi}{d-1}} \mathcal W (I_\lambda,K_\lambda)=
-(1-\xi/2)(\rho')^{-d-1-\frac{2\xi}{d-1}}.}
Using the above obtained expressions of $C_L^+$, $\beta$ and
$\mathcal W$ in Eq.~\eqref{GL1andGL2} we thus have the solutions

\eq{\left\{ \begin{array}{ll} R_{L_<} (z|\rho,\rho') &=
-\frac{(\rho')^{d+1+2\xi/(d-1)}}{1-\xi/2} \left(
h_L^+(\rho)h_L^-(\rho')+\left(\frac{C_L^-}{C_L^+}\right)
h_L^-(\rho)h_L^-(\rho')\right)
\\ \\
R_{L_>}(z|\rho,\rho') &=
-\frac{(\rho')^{d+1+2\xi/(d-1)}}{1-\xi/2}\left(
h_L^-(\rho)h_L^+(\rho')+\left(\frac{C_L^-}{C_L^+}\right)
h_L^-(\rho)h_L^-(\rho')\right)
\end{array}\right.\label{eq:R_L}}
with $C_L^- / C_L^+$ obtained from Eq.~\eqref{cLmatrix}.  We have
calculated in appendix B the asymptotic expression for $C_L^- / C_L^+$
for the two Prandtl number cases $\wp \ll 1$ and $\wp \gg 1$,

\eq{\frac{C_L^-}{C_L^+} = - \frac{\partial h_L^+(1) - \Lambda_L
h_L^+(1)}{\partial h_L^-(1) - \Lambda_L h_L^-(1)}\label{fraktio}}
with leading order contribution to $\Lambda_L$ as

\eq{\Lambda_L = \frac{\partial h_{M}^\pm (1)}{h_{M}^\pm (1)} \ }
with either the $+$ or the $-$ solution understood, the choice
depending on the Prandtl number.

\section{Dynamo effect}

The mean field dynamo effect for the 2-point function of the magnetic
field corresponds to the case when the evolution operator $\mathcal M$
has positive (possibly generalized) eigenvalues.  Eigenvalues
correspond to poles, in $z$, of the resolvent given in
Eq.~\eqref{eq:R_L} and generalized eigenvalues to branch cuts.  The
$z$ dependence is not seen explicitly in Eq.~\eqref{eq:R_L}, but
recall from Eq.~\eqref{hLpm} that the $h_L^\pm$ depend on $z$, and we
see from Eq.~\eqref{fraktio} and matter in Appendix~\ref{sec:frac-C}
that there is further dependence through $C_L^-/C_L^+$.

The $h_L^\pm$ (see Eq.~\eqref{hLpm}) depend on the square root of $z$,
and since the Bessel functions are analytic on the complex right
half-plane, this square root dependence leads directly to a branch cut
along the negative real axis in the $z$ dependence of the resolvent.
This corresponds to a heat equation like continuum spectrum of
decaying modes, these modes don't contribute to the dynamo effect.

Any other possible contributions to the spectrum come from the
fraction $C_L^-/C_L^+$.  An expression for the latter is given in
Eq.~\eqref{fraktio}, with $\Lambda_L$ computed in
Appendix~\ref{sec:frac-C}.
%% Consider explicitly the second term inside the parentheses of
%% Eqs.~\eqref{eq:R_L} which we denote by
%%
%% \eq{ C (z|\rho,\rho'):=\left( \frac{C_L^-}{C_L^+}\right)
%% h_L^-(\rho) h_L^-(\rho'). \label{eq:C-cond}}
%% %
%% %
%% %
%%
%% The expression for the fraction \eqref{fraktio} with $\Lambda_L$
%% computed in Appendix~\ref{sec:frac-C} can then be studied for other
%% contributions to the spectrum.
Equation~\eqref{fraktio} can be simplified by noting that (using
Eqs.~\eqref{hLpm} and \eqref{w})

\eq{\partial h_L^\pm (1) = - \left( \frac{d}{2}+\frac{\xi}{d-1}
\right) \tilde Z_\lambda^\pm (w) +
(1-\xi/2) \partial_w \tilde Z_\lambda^\pm (w).}
Then we can write

\eq{ \frac{C_L^-}{C_L^+}=- \frac{(1-\xi/2) w  I_\lambda' (w)
 - \left[\frac{d}{2}+\frac{\xi}{d-1}+\Lambda_L \right]
I_\lambda(w) }{(1-\xi/2) w  K_\lambda' (w)
 - \left[\frac{d}{2}+\frac{\xi}{d-1}+\Lambda_L \right]
K_\lambda(w)} \ ,\label{eq:C-quot}}
where we have replaced the partial derivative symbols with primes.  We
underline again that $w$ depends on the square root of $z$, so the
complex plane minus the negative real line for $z$ corresponds to the
$\Re w > 0$ half-plane for $w$.  Since Bessel functions are analytical
on this half-plane, the new singularities introduced by $C_L^-/C_L^+$
may come either from singularities of $\Lambda_L$ or zeros of the
denominator $C_L^+$.  The latter (i.e.\ $C_L^+ = 0$) may be written
more conveniently for future needs as

%% We are mainly interested in the \emph{existence} of the dynamo effect
%% so we try to determine if the spectrum has a positive part.  To do
%% this, one only needs to analyze the denominator of the integrand for
%% singularities, i.e. to consider the equation

%% \eq{(1-\xi/2) w K_\lambda ' (w) -
%% \left[\frac{d}{2}+\frac{\xi}{d-1}+\Lambda_L \right] K_\lambda(w)
%% =0 \label{denominator}}
%
%
%% obtained from the denominator of Eq.~\eqref{eq:C-quot}.  Positive
%% solutions $w$ of this equation correspond to a point spectrum and
%% branch cuts to a continuous spectrum.

%% In what follows it will be more convenient to write
%% Eq.~\eqref{denominator} as

\begin{equation}
\label{denominator2}
  w \frac{K'_\lambda (w)}{K_\lambda (w)}
=
  \frac{2}{2-\xi}
  \left[\frac{d}{2} +\frac{\xi}{d-1}+ \Lambda_L \right]
=:
  {\tilde\Lambda}_L
\,,
\end{equation}
where we define ${\tilde\Lambda}_L$.

Finally we remind the reader that $z$ corresponds to the growth rate
with respect to the reduced time $\tau$ rather than real time $t$, and
the real growth rate is thus $(\tau/t)z$, where for the $P\to0$ case
from Eq.~\eqref{eq:adim-small-P} we have $\tau/t = D_1
l_\kappa^{\xi-2}$ and for the $P\to\infty$ case from
Eq.~\eqref{eq:adim-large-P} we have $\tau/t = D_1 l_\nu^{\xi-2}$.

\subsection{Prandtl number $\wp \to 0$}
\label{sec:dynamo:sub:small-P}

Based on asymptotics in Appendix~\ref{sec:frac-C:sub:wp-ll-1}, we
expect that $\Lambda_L$ does not introduce new singularities in the
small Prandtl case, so we only need to deal with
Eq.~\eqref{denominator2}.

Let us first study the large $w$ asymptotics of the two sides of
Eq.~\eqref{denominator2}.  As shown in
Appendix~\ref{sec:frac-C:sub:wp-ll-1}, Eq.~\eqref{lambdaL1}, in the
limit of vanishing Prandtl number and for large $z$, -- or
equivalently large $w$,-- from the medium range solutions of
Eq.~\eqref{largez} we get

\eq{\Lambda_{L} = (1-\xi/2) w \frac{I_{1+d/2}\left( (1-\xi/2) w
\right)}{I_{d/2}\left( (1-\xi/2) w \right)}. \label{eq:Lambda_L-sl}}
For large $w$ we deduce from the asymptotic properties of Bessel
functions \cite{table} that $\Lambda_L \sim (1-\xi/2)w$.  In the same
manner we deduce that the left hand side of Eq.~\eqref{denominator2}
is $w K'_\lambda(w)/K_\lambda(w) \sim -w$.

Next we remark that if $\lambda$ is pure imaginary, which based on the
definition of $\lambda$, Eq.~\eqref{lambda}, happens for

\eq{\xi > \xi^* := (d-1)\left( \sqrt{\frac{d-1}{2(d-2)}}
-\frac{1}{2} \right)\,, \label{def:xi-star}}
then $K_\lambda (w)$ has an infinity of positive zeros (may be seen
from its small $w$ development), accumulating at $w=0$, and more
importantly it has a largest zero (may be seen from its large $w$
asymptotics), which we shall denote by $w_0$.  At each zero of
$K_\lambda (w)$, the l.h.s.\ of Eq.~\eqref{denominator2} has a pole.

Since for large $w$, asymptotically $K_\lambda (w) \sim (\pi/2w)^{1/2}
\exp(-w) > 0$, we must have $K'_\lambda (w_0) > 0$ (note that we can
exclude $K_\lambda (w_0) = K'_\lambda (w_0) = 0$, since the Bessel
function is the solution of a second order homogeneous differential
equation).  Hence the pole of $w K'_\lambda(w)/K_\lambda(w)$ at $w_0$
has a positive coefficient.

\begin{figure}
\begin{center}
\psfrag{w}{$w_0$}
\includegraphics[width=0.7\textwidth]{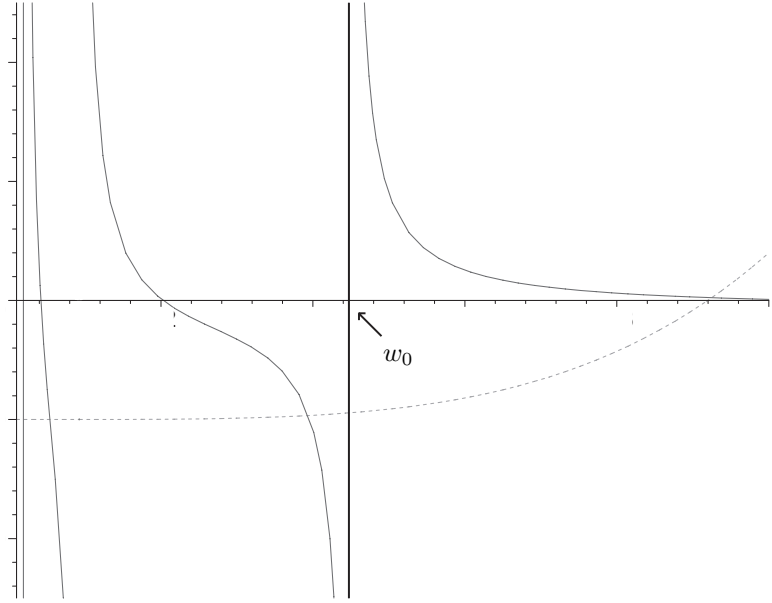}

\end{center}
\caption{Based on the asymptotic properties of the left and right
hand sides of Eq.~\eqref{denominator2}, when $\lambda$ is pure
imaginary and $K_\lambda (w)$ has a largest zero $w_0$, the latter
has to be a lower bound on the largest solution of
Eq.~\eqref{denominator2}. The dashed line depicts the right hand
side of the equation.} \label{fig:lower-bound}\label{FIGzeros}

\end{figure}

From the above asymptotic properties of the two sides of
Eq.~\eqref{denominator2}, in conjunction with the positivity of the
coefficient of the pole at $w_0$, using continuity of the two sides we
deduce -- see also Fig.~\ref{fig:lower-bound} -- that
Eq.~\eqref{denominator2} admits a solution which is at least as large
as the largest zero $w_0$ of $K_\lambda (w)$.

\begin{figure}
\begin{center}
\includegraphics[width=100mm,height=70mm]{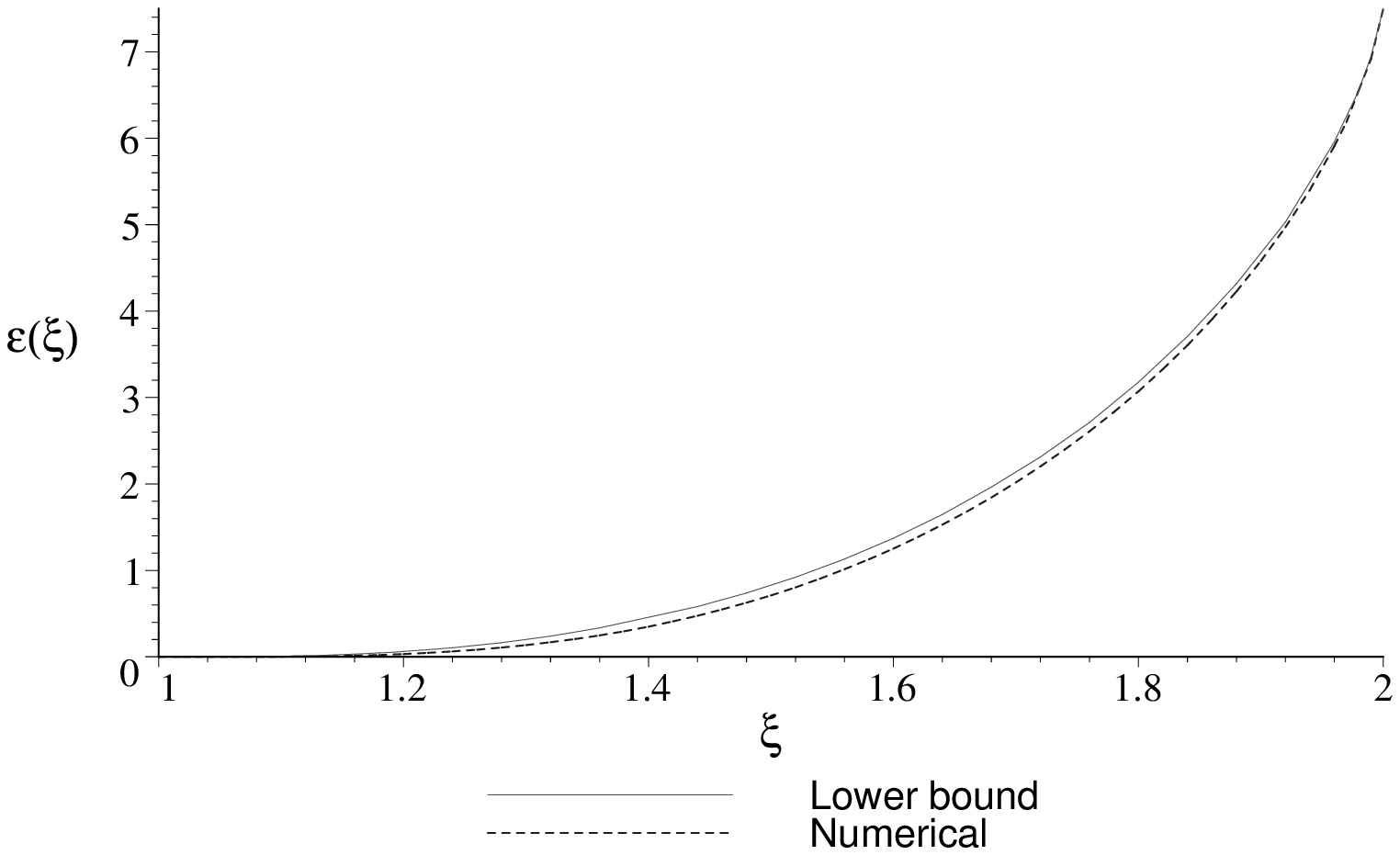}
\includegraphics[width=90mm,height=65mm]{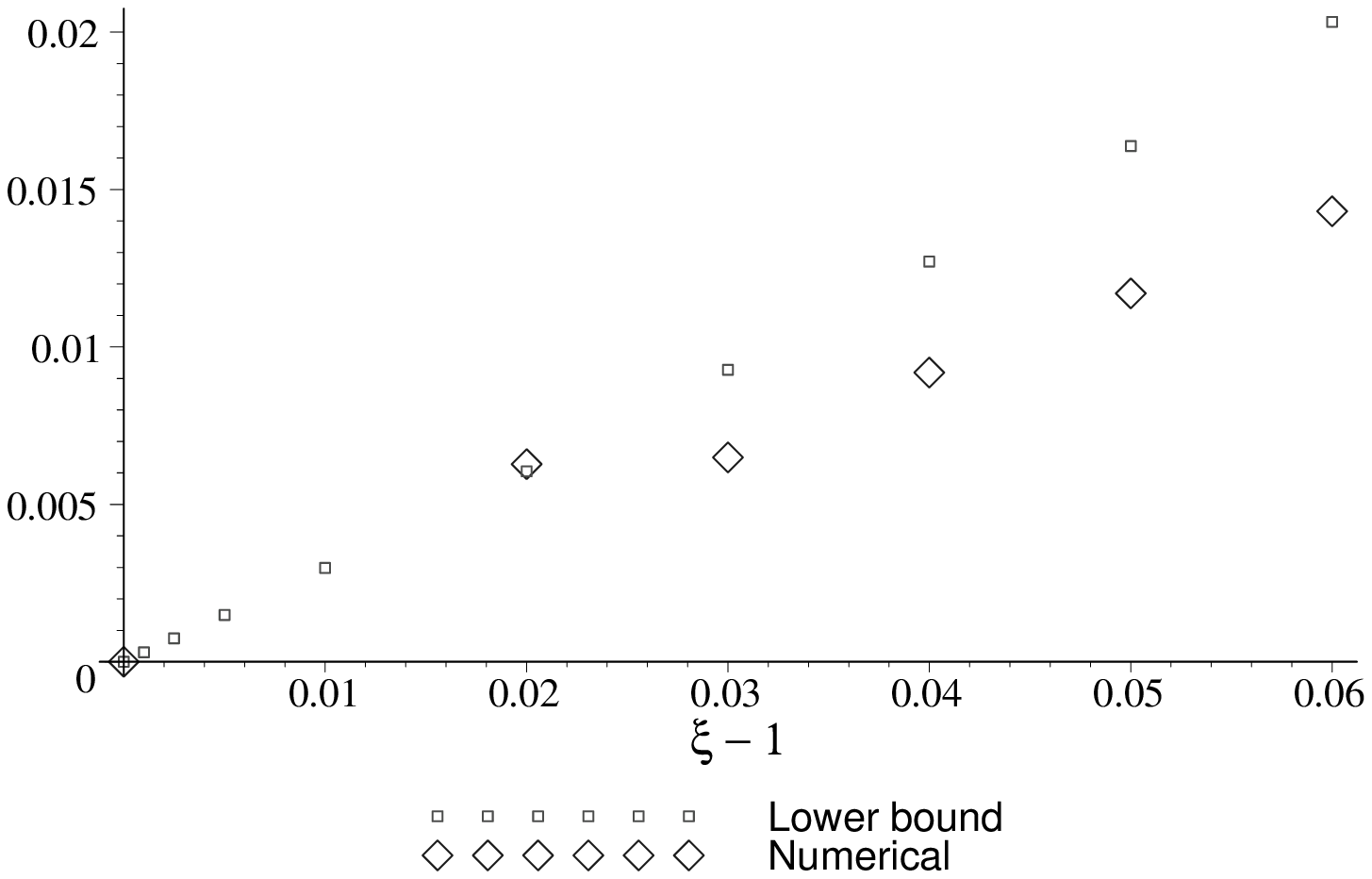}
\includegraphics[width=80mm,height=50mm]{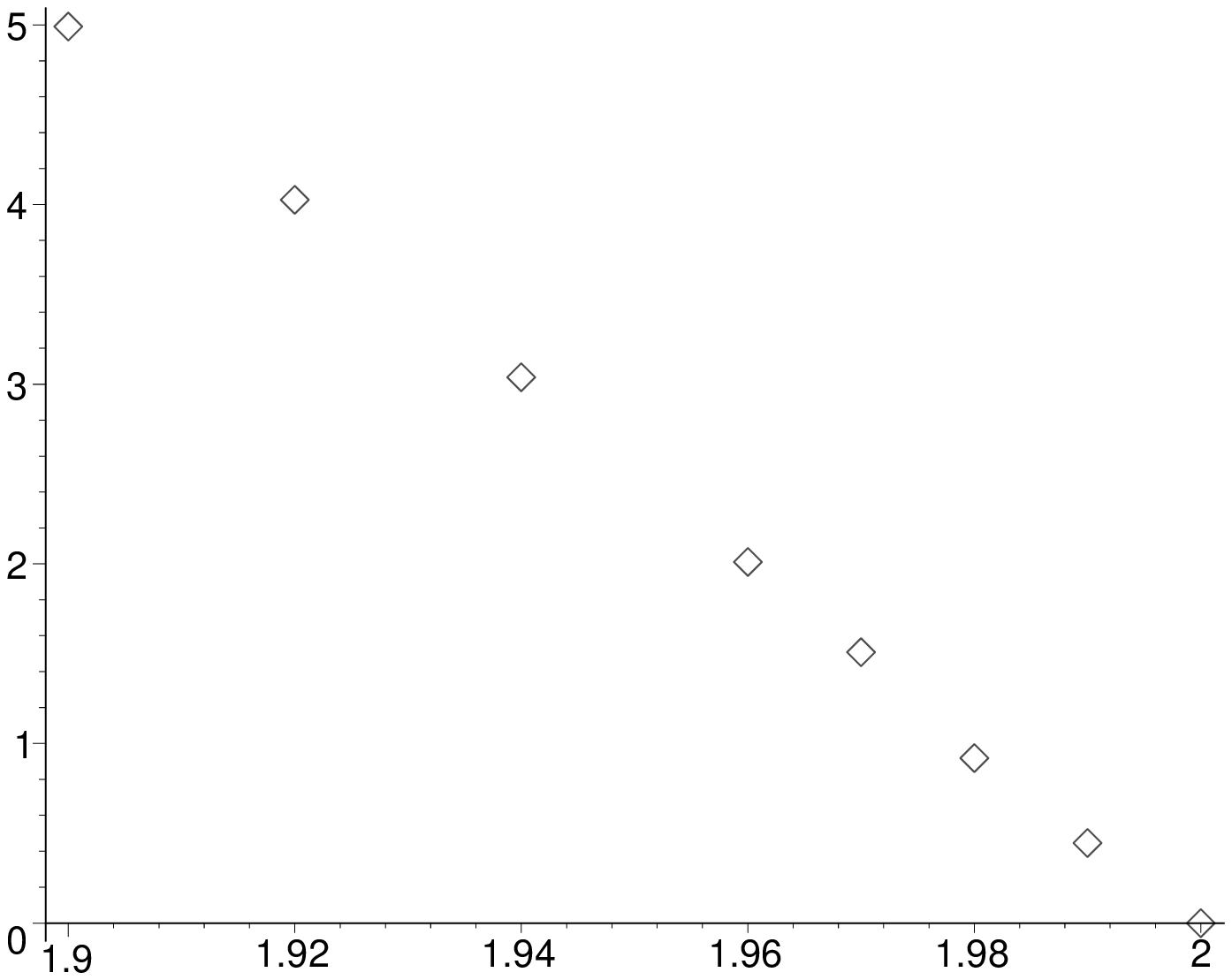}
\end{center}
\caption{In the above figure we have plotted the lower bound $w_0$
with the numerical results of \cite{vincenzi}. The middle figure
shows plots of both data with $1/ \log (\epsilon (\xi))^2$ on the
y-axis. The lower bound data shows linear behavior consistent with
the asymptotics in eq. (\ref{eq:z-xi-star}). We note that there
seems to be a numerical error in the data of \cite{vincenzi} for
$\xi=1.02$. In the lowest figure we have also plotted the
numerical data in \cite{vincenzi} near $\xi=2$ for $(15/2-\epsilon
(\xi))^{3/2}$ showing linear behavior as expected in eq.
\ref{eq:z-xi-2}. The plots in other dimensions look
  similar, except that they begin from the critical value $\xi^* >
  1$.}
\label{FIGepsilon}
\end{figure}

As a convenient means of estimating the solution of
Eq.~\eqref{denominator2}, we may use the lower bound $w_0$.  One
should also be able to obtain an upper estimate, which
%% [based on the asymptotic behaviours discussed above ???],
we expect to be of the same order of magnitude as $w_0$, as argued
below.  For the $d=3$ dimensional case we plot $w_0$ in
Fig.~\ref{FIGepsilon} and it indeed compares well with the numerical
results of \cite{vincenzi}, the latter involving no approximations.
This corroborates our approach.

First we note that, as shown in
Appendix~\ref{sec:Sturm-Liouville:sub:monoton}, $\Lambda_L$ is
increasing.  The small $z$ asymptotics of $\Lambda_L$ is given in
Eq.~\eqref{eq:Lambda_L-ss}.  Two cases are distinguished, depending on
wether ${\tilde\Lambda}_L(z=0) \geq 0$ or not.  If it is then we have
the upper bound $w_0'$, where $w_0'$ is the largest zero of
$K_\lambda'$.  Otherwise we use $(wK_\lambda'(w)/K_\lambda(w))' < -1$
from Appendix~\ref{sec:Sturm-Liouville:sub:Bessel} and get the upper
bound $w_1 = w_0' + |{\tilde\Lambda}_L(z=0)|$.  Fortunately for $\xi$
near $\xi^*$ we have ${\tilde\Lambda}_L(z=0)>0$ (can be seen by
numerical computation for relevant values of $d$), which means that
the $\xi\to\xi^*$ asymptotics computed in
Sect.~\ref{sec:dynamo:sub:asympt} is valid.  The situation for the
$\xi\to2$ asymptotics is more complicated, at the end of
Appendix~\ref{sec:frac-C:sub:wp-ll-1} it is explained why near $\xi=2$
we may use the upper bound $w_0'$: although ${\tilde\Lambda}_L(z=0) <
0$, we can justify ${\tilde\Lambda}_L(z) > 0$ for $z$ the dynamo
growth rate.

%% We therefore expect the expression~\eqref{denominator} to admit
%% solutions analogous to zeros of Bessel functions.

\medskip

We can also study the small $w$ asymptotics of the two sides of
Eq.~\eqref{denominator2}.  As shown in
Appendix~\ref{sec:frac-C:sub:wp-ll-1}, Eq.~\eqref{lambdaL1b}, in the
limit of vanishing Prandtl number and for small $z$, -- or
equivalently small $w$,-- from the medium range solutions of
Eq.~\eqref{smallz} we get

\eq{\Lambda_L = -\xi \sqrt{d/\xi +1} \frac{J_{d/\xi +1} (2
\sqrt{d/\xi +1})}{J_{d/\xi} (2 \sqrt{d/\xi +1})}. \label{eq:Lambda_L-ss}}
which is obviously independent of $w$, so it is in fact the $w\to0$
limit of $\Lambda_L$.
%% and one may verify by a plot that
%% for dimensions 3 to 8 it is positive for $0 < \xi < \xi^*$ ($\xi^*$ is
%% introduced below in Eq.~\eqref{def:xi-star}).
%
Finally we deduce from the asymptotic properties of Bessel functions
\cite{table} that, as $w\to0$, if $\lambda$ is real then the left hand
side of Eq.~\eqref{denominator2} goes to $-\lambda$, and if $\lambda$
is pure imaginary then it has an infinite number of poles at the
positive zeros of $K_\lambda$.

%% The numerical results using the above two alternatives are very
%% similar: as long as the function $\Lambda_L$ stays finite, the
%% presence of zeros is strongly dominated by the zeros of the Bessel
%% functions.  Indeed, by looking at Figure~\ref{FIGzeros}, we can expect
%% the solution to be

%% We have plotted the lower bounds for the growth rates in three [AND
%% FIVE ???]  dimensions in Figure~\ref{FIGepsilon}.

\medskip

So far we have dealt with the case when $\xi^* < \xi \leq 2$, and we
were able to show the existence of a dynamo effect and give a lower
bound on the dynamo growth rate.  We are now going to argue that in
all other situations there is no dynamo effect.

Using the integral representation of the modified Bessel function of
the second kind,

\eq{K_\lambda (w) = \int\limits_0^\infty \!\! dt\, e^{-w \cosh (t)}
\cosh (\lambda t),\label{BesselInt}}
we see that when $\lambda \in \mathbb{R}$ (equivalently when $0 \leq
\xi \leq \xi^*$) and $w>0$, then $K_\lambda(w) > 0$.  On the other
hand using the recurrence relation $K_\lambda'(w) = -K_{\lambda-1}(w)
- \lambda K_\lambda(w)/w$ we get $w K_\lambda'(w) / K_\lambda(w) =
-\lambda -w K_{\lambda-1}(w) / K_\lambda(w) \leq -\lambda$, where the
last inequality follows from the positivity of $K_{\lambda-1}(w)$ and
$K_\lambda(w)$.  Using the power series development of Bessel
functions we also get that $w K_\lambda'(w) / K_\lambda(w) \to
-\lambda$ as $w \to 0$.

%% and $K_\lambda ' <0$, so for $0 \leq \xi \leq
%% \xi^*$ the left hand side of \eqref{denominator2} is negative.  In
%% fact we can also show that it is monotonically decreasing, and we have
%% mentioned above that it starts from $-\lambda$ at $w=0$.

%% we see that when $\lambda \in \mathbb{R}_+$, $K_\lambda > 0$ and
%% $K_\lambda ' <0$, so for $0 \leq \xi \leq \xi^*$ the left hand side of
%% \eqref{denominator2} is negative.  In fact we can also show that it is
%% monotonically decreasing, and we have mentioned above that it starts
%% from $-\lambda$ at $w=0$.

For $3 \leq d \leq 8$ the critical $\xi^*$ takes values between $1$
and $2$, in particular for $d=3$ we get $\xi^*=1$ as expected
\cite{vincenzi, kazantsev, vergassola}.  In this case for $\xi$ such
that $0\leq\xi\leq\xi^*<2$, the right hand side of
Eq.~\eqref{denominator2} is always positive (can be seen
numerically/graphically),
%% and one should also bear in mind that when $\xi$ is far from 2 then we
%% are always in the small $z$ case
so there can be no solutions of Eq.~\eqref{denominator2}, hence no
dynamo.

For $d=2$ one readily verifies that $\xi^* = +\infty$ so there is no
``normal'' dynamo (i.e.\ one with $\xi^* < \xi \leq 2$), and one
further checks that there isn't any ``exceptional'' solution with $\xi
< \xi^*$ (cf.\ Sect.~\ref{sec:remarks:sub:excpt-sols}), by showing
that the left hand side of (\ref{denominator2}) starts from $-\lambda$
at $w=0$ and is decreasing, while the right hand side starts from a
larger value and is increasing.

Finally, for $d \geq 9$ we get $\xi^*>2$ so there is no ``normal''
dynamo either, and plots indicate that the left hand side of
Eq.~\eqref{denominator2} is always negative while its right hand side
is always positive, so again there is no dynamo at all.

%% For $d \geq 9$ we expect no dynamo, since $\xi \leq 2 < \xi^*$ always.

%% the order parameter is $\lambda =
%% (1-\xi/2)^{-1}$, so there is no dynamo in two dimensions.

%% For $\xi> \xi^*$ we have an imaginary $\lambda$ so that the Bessel
%% functions $K_\lambda$ are oscillating on part of the positive axis, as
%% depicted in Fig.~\ref{FIGzeros} and there is always a positive
%% solution and a positive lower bound on it as detailed above.
%% Moreover, both the Bessel function and its derivative have a largest
%% zero (the Bessel function behaves asymptotically as $\propto 1/ \sqrt
%% w \ e^{-w}$), the $K_\lambda '$ one being larger of the two.

We have thus found that in dimensions $3 \leq d \leq 8$ a critical
value $1 \leq \xi^* < 2$ exists above which the dynamo is present and
below which we don't expect it to be present.  In other dimensions we
expect no dynamo for any value of $\xi$.

\subsection{Asymptotics for $\xi$ near $\xi^*$ and 2}
\label{sec:dynamo:sub:asympt}

In this subsection we give estimates of the growth rate of the dynamo
in the cases when $\xi$ is near the critical value $\xi^*$ above which
dynamo is present, and when $\xi$ is near its maximum possible value
2.  What we need is an estimate of the largest solution $w$ of
Eq.~\eqref{denominator2}, from which the corresponding growth rate $z$
is immediately deduced through Eq.~\eqref{w}.  As an order of
magnitude estimate for the solution, the simplest is to take the
largest zero $w$ of $K_\lambda(w)$, as can be seen by inspecting
Figure~\ref{FIGzeros}.

The case $\xi \searrow \xi^*$, corresponding to $\lambda \to 0$ along
the imaginary axis, is somewhat simpler, it can be dealt with starting
from the integral representation \eqref{BesselInt}.  Since
$\xi>\xi^*$, the parameter $\lambda$ is imaginary and we write
$\lambda = i\tilde\lambda$ with $\tilde\lambda \in \mathbb{R}$, hence
$K_{i\tilde\lambda}(w) = \int_0^\infty dt \exp(-w\cosh(t))
\cos(\tilde\lambda t)$.  Now $\cos(\tilde\lambda t)$ is positive near
$t=0$ and it becomes negative for the first time only for $t >
\pi/(2\tilde\lambda)$.  On the other hand the term $\exp(-w\cosh(t))$
is basically a double exponential and decays very fast for $w\cosh(t)
> 1$.  So in order to get for the previous integral a non-positive
result, we need $w_0  \cosh(\pi/(2\tilde\lambda)) \sim 1$ implying
$w_0 \sim \exp(-\pi/(2\tilde\lambda))$.  Through Eq.~\eqref{w} one
deduces the behaviour $\ln z \sim c/\tilde\lambda$, and since
$\tilde\lambda \propto (\xi-\xi^*)^{1/2}$ near $\xi^*$ (the term
under the square root in Eq.~\eqref{lambda} is expected to have a
simple root at $\xi=\xi^*$), we finally have
\begin{equation}
\label{eq:z-xi-star}
  \ln z
\propto
  (\xi-\xi^*)^{-1/2}
\end{equation}
near $\xi^*$.

We now pass to the asymptotics of the case $\xi\to2$.  Under this
limit $\lambda$ diverges as $(2-\xi)^{-1}$, along the complex axis.
From plots or asymptotical formulae we can convience ourselves that
the largest zero of $K_\lambda(w)$ occurs at $w \sim |\lambda|$, so
that is the region where we are going to look for it.  We write $K$ in
terms of the Hankel function of first kind $K_\lambda(w) =
\frac{i\pi}{2} e^{i\pi\lambda/2} H^{(1)}_\lambda(iw)$ and use the
approximation formula in the transitory region (i.e.\ when parameter
and argument of the Bessel function are of same order):
$$
  H^{(1)}_\lambda(\lambda+t\lambda^{1/3})
\sim
  \frac{\sqrt{2}e^{i\pi/6}} {3\lambda^{1/3}t}
  H^{(1)}_{1/3}(\frac{\sqrt{8}}{3} t^{3/2})
\,.
$$
We deduce that there is some $t_0 > 0$ such that $w_0 \sim |\lambda| -
t_0 |\lambda|^{1/3} = |\lambda|(1 - t_0|\lambda|^{-2/3})$.  Combining
Eqs.~\eqref{w} and \eqref{lambda} we get
\begin{equation}
\label{eq:z-xi-2}
  z
\approx
z_2 - c(2-\xi)^{2/3}
\,,
\end{equation}
valid for $\xi$ near 2, where $c>0$ is some constant of order unity
and $z_2$ was introduced in Eq.~\eqref{def:z_2}.

\subsection{Prandtl number $\wp \to \infty$}

For large Prandtl number the analysis proceeds exactly as in the
previous section, except that we have a different $\Lambda_L$.  From
Eqs.~\eqref{lambdaL2} and \eqref{zeta} we have, using the definition
of $\zeta$ from Eq.~\eqref{def:zeta},

\eq{\Lambda_L &= \zeta -\frac{2}{d-1}-\frac{d}{2}\,.
\label{eq:Lambda_L-l}}
%
%% \eq{\Lambda_L &= \zeta -\frac{2}{d-1}-\frac{d}{2} =
%% \frac{\sqrt{d(d^3-10d^2+9d+16) +4(d-1) z}}{2(d-1)}
%% -\frac{2}{d-1}-\frac{d}{2} \nonumber \\ &= \sqrt{\frac{d(d^3 - 10
%% d^2 + 9 d +16)}{4 (d-1)^2} + (1-\xi/2)^2 w^2}
%% -\frac{2}{d-1}-\frac{d}{2}\,. \label{eq:Lambda_L-l}}
%
%
%
%
There is now a branch cut originating from $z_2$ (defined in
Eq.~\eqref{def:z_2})
%\eq{z_2= -\frac{d}{4(d-1)} (d^3-10d^2+9d+16) }
%
%
extending to infinity along the negative real axis.  When $3 \leq d
\leq 8$, $z_2$ is positive and the branch cut extends up to the
positive value $z_2$ along the positive real axis, i.e.\ the spectrum
has a continuous positive part for \emph{all} $\xi \in [0,2]$.
Another major difference when comparing to the small Prandtl number
case is that the spectrum is continuous also in the positive part.

We conclude that the dynamo is present for all $\xi$ for large Prandtl
numbers.

\section{Some remarks}
\label{sec:remarks}

\subsection{Schr\"odinger operator formalism}

We have performed our analysis in the diffusion process setting, but
it could have equally well been done in the Schr\"odinger operator
formalism (e.q.\ \cite{vincenzi}), basically with the same kind of
cutting up and piecing together technique.

Consider the zero energy Schrödinger equation

\begin{figure}
\begin{center}
\includegraphics[width=0.7\textwidth]{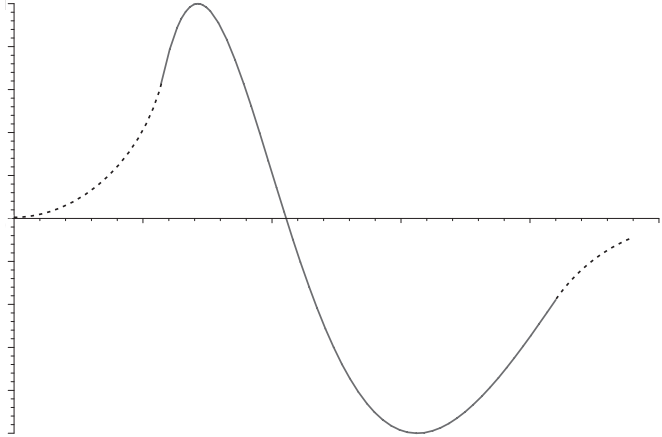}

\end{center}
\caption{Sketchy plot of the medium range zero energy solution
$\sin(\sqrt 15 /2 \log(r))$ crossing zero, implying the existence
of a negative energy state. The dashed lines correspond to regions
where the middle range approximation is no longer valid. As the
range grows when increasing the Prandtl number, more zeros will
emerge.} \label{zeros}
\end{figure}

\eq{\psi''(r) = V(r) \psi(r),}
where $V(r)=m(r) U(r)$ is the effective potential. The potential
$V$ behaves as $2/r^2$ at very short and long scales, but as
$-4/r^2$ at the medium range.  The medium range solutions are
$\sin (\sqrt 15/2 \log (r))$ and $\cos (\sqrt 15/2 \log (r))$.
When the Prandtl number is increased, the medium range region is
stretched, and it is clear that for sufficiently large Prandtl
numbers the solutions cross zero an increasing number of times
(see Fig.~\ref{zeros}).  According to a well known theorem, such a
solution cannot be a ground state (see e.g. \cite{reedsimon}
pp.~90). In fact, the number of zeros of the solution (with
nonzero derivative and excluding the zero at $r=0$) is the number
of negative energy states, which implies the existence of
unbounded growth.

\subsection{Finite magnetic Reynolds number effects}

Let us finally touch upon some questions not discussed in the text.
Our method allows us in principle, without further complications, to
estimate the critical magnetic Reynolds number (dependent on velocity
roughness exponent $\xi$ and space dimension $d$) at which dynamo
effect sets in, and the growth of the dynamo exponent with Reynolds
number.  However we get only a logarithmic estimate whose uncertainty
is at least an order of magnitude or even two, which makes it not too
useful.  Notwithstanding, we would like to mention that the estimates
we would obtain this way are hardly compatible with numerical results
of \cite{vincenzi}, our thresholds being significantly lower.  This
issue is currently clarified with D.~Vincenzi.

\subsection{Exceptional solutions}
\label{sec:remarks:sub:excpt-sols}

An other issue is that of the existence of ``exceptional'' dynamos.
It seems to us that the ``typical'' dynamo (note that we consider here
only the infinte magnetic Reynolds number case) corresponds to the
situation when our $\xi>\xi^*$, in which case there is an infinite
discrete spectrum of growing modes.  However our equations do not
exclude {\em a priori} the possibility of a single growing mode at
some $\xi<\xi^*$.  In fact, if we take for example, at a formal level,
$d=2.125$ then $\xi^* \approx 1.82$ and for $\xi' \leq \xi < \xi^*$
(where $\xi'$ is some value of which we only need to know here that
$\xi' < 1.77$), Eq.~\eqref{denominator2} will have, in what we have
called the small $z$ approximation (cf.\ Eq.~\eqref{ineq:z-P-small}),
a single solution $w_0 > 0$.  If we take $\xi=1.77$ then $w_0 \approx
0.077$ and $z \approx 8.8 \cdot 10^{-5} \ll 1$ in a self-consistent
manner.  However it remains to be known if such a solution is not just
an artefact of our resolution method, and if not, then to see if one
can construct a model where such solutions occur for the more physical
value of $d=3$.

\section{Conclusions}

The mean-field dynamo problem was considered in arbitrary space
dimensions.  We have shown that, to obtain the spectrum of the dynamo
problem, the equation \eqref{denominator2} has to be solved for $w$,
from which the growth rate $z$ can be expressed through Eq.~\eqref{w}.
The quantity $\Lambda_L$ appearing in Eq.~\eqref{denominator2} is
given, for small magnetic Prandtl numbers, by either
Eq.~\eqref{eq:Lambda_L-sl} or \eqref{eq:Lambda_L-ss}, depending on
which of the self-consitent  conditions \eqref{ineq:z-P-large} or
\eqref{ineq:z-P-small} is verified (note that this leaves a gap
between with no formula).  For large magnetic Prandtl number we have
to use Eq.~\eqref{eq:Lambda_L-l} instead.

It was observed that, in our model, the dynamo can only exist when $3
\leq d \leq 8$.  The results for small Prandtl numbers were shown to
confirm previous results \cite{vincenzi,kazantsev} obtained in three
dimensions.  For $d > 3$ a critical value for $\xi$ was found, above
which the dynamo is present, which is larger than the three
dimensional critical value $\xi^*=1$.  Furthermore, in the vanishing
Prandlt number limit we have obtained the asymptotic
estimates~\eqref{eq:z-xi-star} and \eqref{eq:z-xi-2}, which are in
good qualitative agreement with numerical simulations of
\cite{vincenzi}.

For large Prandtl numbers it was shown that the dynamo exists for
\emph{all} $\xi$ and that the spectrum is continuous.  We hope our
work will contribute to clarifying this somewhat controversial issue.
The physical idea behind our explanations is that at large magnetic
Prandtl number the magnetic field can feel the smooth scales of the
fluid flow (they are not ``wiped out'' by magnetic diffusivity), and
correlations in the velocity field above the magnetic diffusive scale
$l_\kappa$ won't do more harm to the dynamo than if we had a Batchelor
type flow with no correlations of velocity at scales significantly
larger than $l_\kappa$.

Our methods were based on approximating piecewise the evolution
operator of the two point function of the magnetic field.  This
approximation introduces inaccuracies and one may ask how these
influences the fine details of our reasoning, which relied on not so
evident estimates.  We think that the general picture sketched up
should be valid for the exact problem also, based on the good
agreement with available numerical data from the litterature.  Since
for $\xi=\xi^*$ and $\xi=2$ one can find the fastest growing mode
explicitly, it should also be possible to do a perturbation theory
around these points and place our results on a firmer ground.  This is
however left for future work.

\section{Acknowledgements}

H.A. would like to thank P. Muratore-Ginanneschi and A. Kupiainen
for useful comments, suggestions and discussions related to the
problem at hand. P.H. would like to thank A. Kupiainen for
inviting him to work at the Mathematics Department of Helsinki
University, where most of this work has been done. The work of
H.A. was partly supported by the Vilho, Yrjö and Kalle Väisälä
foundation. We would also like to thank Dario Vincenzi for
discussing his results with us and providing data from his
simulatons.

\newpage
\appendix
\section{PDE for the 2-point function of $\maB$}
\label{apx:PDE-2-B}

We rewrite equation \eqref{equation} as an It\^o type SPDE (following
the formalism of \cite{kupiainen2}, sect. 5)

\eq{ dB_i + d \maw \cdot \nabla B_i - \maB \cdot \nabla dw_i -
\kappa' \Delta B_i dt = 0,}
where $\kappa'=\kappa + D /2$ with $D$ defined as

\eq{D_{ij}(0) = D \delta_{ij}.}
The new diffusion term in $\kappa'$ emerges by advecting the
magnetic field along the particle trajectories similarly as in the
passive scalar case by using the It\^o formula. It will cancel out
eventually, as it should. We can express the above equation more
conveniently by defining

\eq{d b_i (t,x) = -\mathcal D_{ijk}^x \left( B_j(t,x) d w_k (t,x)
\right),}
where $\mathcal D_{ijk}^x = \delta_{ij} \partial_k^x -
\delta_{ik}\partial_j^x$ .\footnote{This is just a rewriting of
the expression $\nabla \times (\maB \times \mav)$ for
incompressible fields $\maB$ and $\mav$} The equation is then
simply

\eq{dB_i  - \kappa' \Delta B_i dt = d b_i.}
For a function $F$ of fields $\maB$, we have the (generalized)
It\^o formula,

\eq{d F\left( \maB(t,\cdot) \right) = \inte d x \frac{\delta
F}{\delta B_i(x)} \left[ \kappa' \Delta B_i dt +db_i \right] +
\frac{1}{2}\inte dx dy \frac{\delta^2 F}{\delta B_i (x)\delta B_j
(y)} \mathbf{E} \left(db_i(t,x) db_j (t,y)\right).}
The advecting velocity field is a time derivative of a Brownian
motion on some state space, that is

\eq{\mathbf{E} dw_i (t,x) dw_j (t,y) = dt D_{ij}(x-y) ,}
where $D_{ij}$ was defined in Eq.~\eqref{correlation}.  This means
that

\eq{\mathbf E db_i (t,x) db_j (t,y)= \mathcal D_{ikl}^x \mathcal
D_{jmn}^y \left( u_k (t,x) u_m(t,y) D_{ln}(x-y) \right).}
We apply this to $F= u_i(t,x)u_j (t,y)$, denote $G_{ij}(x-y) = \mathbf
E u_i(t,x) u_j (t,y)$ and use the decomposition $D_{ij}(x-y)=D
\delta_{ij}-d_{ij}(x-y)$ introduced in Eq.~\eqref{structure}.  Noting
that terms proportional to $dw$ disappear, we obtain the equation for
the two point function:

\eq{\frac{d }{dt}G_{ij}(t,\mar) = 2 \kappa \Delta G_{ij}(t,\mar) +
\mathcal F_{ij} ,\label{equation2}}
where $\mar = x-y$ and

\eq{ \mathcal F_{ij} = d_{\alpha\beta}G_{ij,\alpha\beta}-d_{\alpha
j,\beta} G_{i\beta,\alpha} - d_{i \beta,\alpha} G_{\alpha j,\beta}
+ d_{ij,\alpha\beta} G_{\alpha\beta} \label{F}}
where the indices after commas denote partial derivatives. Note that
this depends only on $\kappa$, not $\kappa'$, i.e. the constant part
$D \delta_{ij}$ of the structure function is absent.  Using the
decomposition \eqref{symmetry} and the explicit form of the long
distance velocity structure function \eqref{longstructure} we get from
Eq.~\eqref{equation2} two equations for $G_1$ and $G_2$,

\eq{  \frac{dG_1}{dt} = \frac{2 \kappa}{r^2} \left(2G_2+ (d-1) r
G_1'+r^2 G_1''\right) +\mathcal A ,\label{A9}}
and

\eq{ \frac{dG_2}{dt} = \frac{2 \kappa}{r^2} \left( -2 d G_2 +
(d-1) r G_2' + r^2 G_2'' \right) + \mathcal B .\label{A10}}
The symbols $\mathcal A$ and $\mathcal B$ are the terms arising
from the interaction with the (long distance) velocity fields.
Using the relations \eqref{G1G2} for $G_1$ and $G_2$ in terms of
$H$, their explicit form is as follows:

\eq{\frac{\mathcal A}{D_1 r^{-2+\xi}} &=  \xi (d-1)(d-3+\xi)H(r)
\nonumber \\ \ldots &+ (2-d-2d^2+d^3+(-5+d+2d^2)\xi +(1+d)\xi^2)r
H'(r) \nonumber \\ \ldots &+ (2d(d-1)+(d+1)\xi))r^2 H''(r)
+(d-1)r^3H'''(r),}
\eq{- \frac{\mathcal B}{D_1 r^{-2+\xi}} &= -\xi(d-1)(2-\xi)H(r)
\nonumber \\ \ldots &+ ((1-d^2)+(d-5+2d^2)\xi + 4\xi^2-\xi^3)r
H'(r) \nonumber \\ \ldots &+ (d+1)(d-1+\xi)r^2 H''(r) + (d-1)r^3
H'''(r) .}
Now we can just add the equations \eqref{A9} and \eqref{A10}, and
by using $G_1 + G_2 = (d-1)H$ we get

\eq{\frac{d H}{dt} &= \xi (d-1)(d+\xi) D_1 r^{-2+\xi}H(r)
\nonumber \\ \ldots &+ \left( 2(d+1)\kappa +  (d^2 -1 +
2\xi)D_1 r^\xi \right) r^{-1}H'(r)  \nonumber \\
\ldots &+  (2 \kappa +(d-1)D_1 r^\xi)H''(r) . }\\
\newpage
\
\section{Computation of the fraction $C_L^- / C_L^+$}
\label{sec:frac-C}

By evaluating the matrix multiplications on the right hand side of
Eq.~\eqref{cLmatrix}, we can write the fraction $C_L^- / C_L^+$ as

\eq{ \frac{C_L^-}{C_L^+}=- \frac{\partial h_L^+ (1) - \Lambda_{L}
h_L^+ (1)}{\partial h_L^- (1) - \Lambda_{L} h_L^- (1)},}
where $\Lambda_L$ can be written as the following nested
expression:

\eq{\left\{ \begin{array}{ll} \Lambda_{L} &=\frac{\partial h_{M}^+
(1) + \Lambda_{M} \partial h_{M}^- (1)}{ h_{M}^+ (1) + \Lambda_{M}
h_{M}^- (1)}
\\ \Lambda_{M} &=- \frac{\partial
h_{M}^+ (a_i) - \Lambda_{S} h_{M}^+ (a_i)}{
\partial h_{M}^- (a_i) - \Lambda_{S} h_{M}^- (a_i)}
\\ \Lambda_{S} &= \frac{\partial h_S (a_i)}{h_S (a_i)} .
\end{array} \right.\label{Lambdas}}
This follows from defining

\eq{ \begin{pmatrix} h_S
\\ \partial h_S \end{pmatrix}_{a_i} = h_S(a_i) \begin{pmatrix} 1
\\ \Lambda_{S} \end{pmatrix}}
and writing equation \eqref{cLmatrix} as

\eq{\begin{pmatrix} C_L^+
\\ C_L^- \end{pmatrix} = c \begin{pmatrix}
\partial h_L^- &\!\!\! -h_L^- \\
-\partial h_L^+ &\!\!\!h_L^+
\end{pmatrix}_{1} \begin{pmatrix}
h_{M}^+ &\!\!\! h_{M}^- \\
\partial h_{M}^+ &\!\!\! \partial h_{M}^-
\end{pmatrix}_{1} \begin{pmatrix}
\partial h_{M}^- &\!\!\! -h_{M}^- \\
-\partial h_{M}^+ &\!\!\! h_{M}^+
\end{pmatrix}_{a_i} \begin{pmatrix} 1
\\ \Lambda_{S} \end{pmatrix},}
where $h_S(a_i)$ is absorbed in the coefficient. We have defined
above a constant $c$ which gets cancelled in the end of
computations. It will be used below as well as a generic constant
that does not affect the final results. Multiplying the last
matrix with the vector, we define similarly

\eq{\begin{pmatrix}
\partial h_{M}^- &\!\!\! -h_{M}^- \\
-\partial h_{M}^+ &\!\!\! h_{M}^+
\end{pmatrix}_{a_i} \begin{pmatrix} 1
\\ \Lambda_{S} \end{pmatrix} = \begin{pmatrix} \partial
h_{M}^- - \Lambda_{S} h_{M}^-
\\ - \partial h_{M}^+ + \Lambda_{S} h_{M}^+ \end{pmatrix}_{a_i} = c \begin{pmatrix}
1
\\ \Lambda_{M} \end{pmatrix},}
that is,

\eq{
\label{eq:Lambda_M}
\Lambda_{M} = - \frac{\partial h_{M}^+(a_i) - \Lambda_{S}
h_{M}^+ (a_i)}{\partial h_{M}^- (a_i) - \Lambda_{S} h_{M}^-
(a_i)}}
Doing this again for the second matrix, we obtain similarly

\eq{\Lambda_{L} = \frac{\partial h_{M}^+ (1) + \Lambda_{M}
\partial h_{M}^- (1)}{ h_{M}^+ (1) + \Lambda_{M}
h_{M}^- (1)}}
and finally

\eq{\frac{C_L^-}{C_L^+}=- \frac{\partial h_L^+ (1) - \Lambda_{L}
h_L^+ (1)}{\partial h_L^- (1) - \Lambda_{L} h_L^- (1)}.}

We are interested in the leading order behavior of the fraction
$C_L^- / C_L^+$ only, so we need to determine what happens to
$\Lambda_M$ as $\wp$ approaches zero or infinity. It turns out
that either $\Lambda_M \to 0$ or $\Lambda_M \to \pm \infty$, so to
the leading order,

\eq{\Lambda_{L} = \frac{\partial h_{M}^\pm (1)}{ h_{M}^\pm (1) }\
.}

\subsection{$\wp \ll 1$}
\label{sec:frac-C:sub:wp-ll-1}

Below the suspension dots denote higher order terms in powers of $P$
(or $P^{-1}$ for large Prandtl numbers). Recall from
Eqs.~\eqref{aANDb} and \eqref{smallprnumber} that
\eq{a_1 = l_\nu / l_\kappa = \left(
\frac{d-1}{2}\wp\right)^{1/\xi}.}
The short range solution was

\eq{h_S (\rho) = \rho^{-d/2} \ I_{d/2} \left( \alpha \rho \right)
}
with a temporary notation $\alpha = \sqrt{(z+2
\wp^{1-2/\xi}(2-d-d^2))/(d-1)}$ and note that $|\alpha|$ behaves as
$\wp^{1/2-1/\xi}$.  Using standard relations of Bessel functions
\cite{table} and using the definition for $\Lambda_S$ in
Eq.~\eqref{Lambdas}, we have

\eq{\Lambda_{S} = \frac{\partial h_S (a_1)}{h_S(a_1)} = \alpha
\frac{I_{1+d/2} \left( \left( \frac{d-1}{2}\wp \right)^{1/\xi}
\alpha \right)}{I_{d/2} \left( \left(\frac{d-1}{2}\wp
\right)^{1/\xi} \alpha \right)}.}
Since $\wp$ is small and the arguments of the Bessel functions above
scale as $\wp^{1/2}$, we can use the expansion

\eq{I_{d/2} (u) = u^{d/2} \left( \frac{2^{-{d/2}}}{\Gamma(1+d/2)}
+ \frac{2^{-2-d/2}}{\Gamma(2+d/2)} u^2 + \mathcal O (u^4) \right)}
(and a corresponding one when the order parameter is $1+d/2$) to
conclude that

\eq{\Lambda_{S} = c \wp^{1-1/\xi} + \ldots }
The medium range solutions in the large $z$ case, Eq.~\eqref{largez},
are

\eq{h_{M,1}^{\pm} (\rho) = \rho^{-d/2} \left\{
\begin{array}{cc} I_{d/2} & \\ K_{d/2}  & \end{array}\right.
 \!\!\!\!\!\!\!\!\! \left( \sqrt{\beta} \rho
\right),}
with $\beta = z/(d-1)$. The leading order behavior is

\eq{
\left\{ \begin{array}{ll} h_M^+ (a_1) &= c  + \ldots \\
\partial h_M^+ (a_1) &= c \wp^{ 1/\xi} +
\ldots\\
h_M^- (a_1) &= c P^{-d/\xi}  + \ldots\\
\partial h_M^- (a_1) &= c \wp^{ -1/\xi-d/\xi} +
\ldots
\end{array} \right.
\label{eq:asympt-med-large-z} }
Using these on $\Lambda_M$ as given by Eq.~\eqref{eq:Lambda_M}, we see
that to leading order

\eq{\label{eq:Lambda_M-small-P}
\Lambda_M = c P^{d/\xi +1},}

which goes to zero. Therefore we have

\eq{\Lambda_{L} \sim \frac{\partial h_{M}^+ (1)}{ h_{M}^+ (1)} =
\sqrt{\frac{z}{d-1}} \frac{I_{1+d/2} \left( \sqrt{\frac{z}{d-1}}
\right)}{I_{d/2} \left( \sqrt{\frac{z}{d-1}}
\right)}\label{lambdaL1}.}
Note that, notwithstanding the fractional powers appearing above,
$\Lambda_L$ is a single valued function, indeed near $z=0$ it behaves
as $\Lambda_L \approx z/(d-1)$.  One also notes that in the large $z$
case $\Lambda_L$ is always positive, since the Bessel functions $I$
are positive for positive parameter and argument.

We may perform a similar analysis for the small $z$ approximation,
based on Eq.~\eqref{smallz},

\eq{h_{M,1}^{\pm} (\rho) = \rho^{-d/2} \left\{
\begin{array}{cc} J_{d/\xi} & \\ Y_{d/\xi}  & \end{array}\right.
 \!\!\!\!\!\!\!\!\! \left( \gamma \rho^{\xi/2}
\right),}
with $\gamma = 2\sqrt{d/\xi+1}$. The leading order behavior is

\eq{
\left\{ \begin{array}{ll} h_M^+ (a_1) &= c  + \ldots \\
\partial h_M^+ (a_1) &= c \wp^{ 1-1/\xi} +
\ldots\\
h_M^- (a_1) &= c P^{-d/\xi}  + \ldots\\
\partial h_M^- (a_1) &= c \wp^{ -1/\xi-d/\xi} +
\ldots
\end{array}
\right.
\label{eq:asympt-med-small-z} }
Using these on $\Lambda_M$ (cf.\ Eq.~\eqref{eq:Lambda_M}), we see that
once again $\Lambda_M$ behaves at leading order as given in
Eq.~\eqref{eq:Lambda_M-small-P}, meaning that it goes to zero as $P$
goes to zero.  Therefore we have
\eq{
\label{lambdaL1b}
  \Lambda_L
\sim
  \frac{\partial h_{M}^+ (1)}{ h_{M}^+ (1)}
=
  -\xi \sqrt{d/\xi +1}
  \frac{J_{d/\xi +1} (2 \sqrt{d/\xi +1})}
       {J_{d/\xi} (2 \sqrt{d/\xi +1})}
\,.}

In the particular case of $\xi=2$ the medium range solution can be
explicitly calculated for any $z$, and we have

\eq{h_{M,1}^{\pm} (\rho) = \rho^{-d/2} \left\{
\begin{array}{cc} J_{d/2} & \\ Y_{d/2}  & \end{array}\right.
 \!\!\!\!\!\!\!\!\! \left( \sqrt{\beta} \rho \right)\,,}
where now $\beta=2(d+2)-z/(d-1)$.  The approximations in
Eq.~\eqref{eq:asympt-med-large-z} or \eqref{eq:asympt-med-small-z}
(for $\xi=2$ those two coincide) are valid unfiromly as $\xi$ goes to
2, so when $\beta$ is of order unity, the leading order behaviour
Eq.~\eqref{eq:Lambda_M-small-P} is valid.  Note that for $z=z_2$ we
indeed have $\beta$ of order unity.  Now one deduces that $\Lambda_L =
-\sqrt{\beta} J_{d/2+1}(\sqrt{\beta})/J_{d/2}(\sqrt{\beta})$, and for
$z=z_2$ one finds $\sqrt{\beta} = (d^2-d+4)/4/(d-1)$.  One can then
verify numerically that for $\xi=2$ and $z=z_2$ and relevant values of
$d$ (between 3 and 8 inclusive) we have $\tilde\Lambda_L > 0$.  By
continuity, positivity carries over to values of $\xi$ close to 2 and
the corresponding dynamo growth rate $z$.  This permits us to use near
$\xi=2$ the upper bound $w_0'$ on the largest solution of
Eq.~\eqref{denominator2}, and obtain the asymptotic behaviour of
Sect.~\ref{sec:dynamo:sub:asympt}.

\subsection{$\wp \gg 1$}

Now we have $a_2 = \left((d-1)\wp/2 \right)^{-1/2}$. The short range
solution is in this case

\eq{h_S (\rho) = \rho^{-d/2} \ I_{d/2} \left(\sqrt \wp \alpha'
\rho \right)}
with

\eq{\alpha' = \frac{1}{\sqrt 2} \sqrt{z+2 (2-d-d^2)}.}
Similarly to the $\wp \ll 1$ case,

\eq{\Lambda_S = \sqrt \wp \alpha' \frac{I_{1+d/2} (\sqrt
\frac{2}{d-1} \alpha')}{I_{d/2} (\sqrt \frac{2}{d-1} \alpha'))} =
c \sqrt \wp  + \ldots}
The medium range solutions are now power laws,

\eq{h_M^\pm = \rho^{-d/2-2/(d-1) \pm \delta},}
where

\eq{\delta = \frac{\sqrt{d(d^3-10d^2+9d+16) +4(d-1)
z}}{2(d-1)}\label{zeta}.}
Since $\partial h_M^\pm (a_2) \propto \sqrt \wp h_M^\pm (a_2)$,

\eq{\partial h_M^\pm (a_2)-\Lambda_S h_M^\pm (a_2) = c \sqrt \wp
h_M^\pm (a_2) + \ldots,}
that is,

\eq{\Lambda_M = c \frac{h_M^+(a_2)}{h_M^- (a_2)}+ \ldots = c
\frac{1}{\wp} + \ldots.}
This goes to zero as $\wp \to \infty$, and we have

\eq{\Lambda_L \to \frac{\partial h_M^+ (1)}{ h_M^+ (1)} = \zeta -
\frac{2}{d-1} - \frac{d}{2} \label{lambdaL2}.}
where $\zeta$ was defined in Eq.~\eqref{def:zeta}.  In fact we
wouldn't have needed to worry if the limit of $\Lambda_M$ was infinite
or zero. The difference would only be a different sign of $\zeta$,
which doesn't affect anything since it is the presence of the branch
cut alone which determines the positive part of the spectrum.

\section{Some Sturm-Liouville theory}
\label{sec:Sturm-Liouville}

Consider the following general second order linear eigenvalue problem,
where $a,b,c$ are positive functions and $z\in\mathbb{R}$:
\begin{equation}
\label{eq:gen-2o}
  a(\rho)h''(\rho) + b(\rho)h'(\rho) + c(\rho)h(\rho)
=
  z h(\rho)
\,.
\end{equation}
Introduce $g=h'/h$, then $g$ verifies the first order non-linear
(Riccati) differential equation
\begin{equation}
\label{eq:g-prime}
  g'
=
  \frac{z-c-bg-ag^2}{a}
\,.
\end{equation}
Note that a zero of $h$ corresponds to a pole of $g$, and the pole is
always such that as $\rho$ increases $g$ goes to $-\infty$ and comes
back at $+\infty$ (since if $h$ is positive before crossing zero then
its derivative must be negative and vice versa).

\subsection{Montonicity of solutions in $z$}
\label{sec:Sturm-Liouville:sub:monoton}

Consider for Eq.~\eqref{eq:gen-2o} the initial condition $h'(0)=0$ and
$h(0)>0$ which in particular implies $g(0)=0$.  Now consider
Eqs.~\eqref{eq:gen-2o} and \eqref{eq:g-prime} for two different values
of $z$, say $z_1$ and $z_2$, and denote the corresponding solutions by
$h_1$, $g_1$ and $h_2$, $g_2$ respectively.  We show that if $z_1 >
z_2$, then $g_1(\rho) > g_2(\rho)$ for $\rho$ less than the first zero
of $h_2$.

This can be seen as follows.  First, the assertion is true near
$\rho=0$ since $g_1'(0) = (z_1-c(0))/a(0) > (z_2-c(0))/a(0) = g_2'(0)$
while $g_1(0)=g_2(0)=0$.  Now suppose that at some point the ordering
of $g_1$ and $g_2$ changes, this means that the two have to cross,
i.e.\ for some $\rho$ we have $g_1(\rho) = g_2(\rho) = G$.  However at
this point $g_1'(\rho) = (z_1 - c(\rho) - b(\rho)G -
a(\rho)G^2)/a(\rho) > (z_2 - c(\rho) - b(\rho)G - a(\rho)G^2)/a(\rho)
= g_2'(\rho)$, meaning that $g_1$ cannot cross $g_2$ downwards, which
is a contradiction.

\subsubsection*{Application 1}

From the above it also follows that the first zero of $h_1$ is larger
than the first zero of $h_2$.  Indeed $h_2$ has no zero before its
first zero, so $g_2$ doesn't go to $-\infty$ before that point,
implying that $g_1$ neither since $g_1>g_2$, thus $h_1$ has no zero
either before the first zero of $h_2$.

A particularly useful application of this is to use the position of
the first zero of the solution with $z=0$ as a lower bound on the
first zero of any solution for $z>0$.

%% Apply the above with $z_2=0$.  In particular if $h_2$ has no zero then
%% $g_2$ never goes to $-\infty$, implying that $g_1$ neither, which
%% implies that $g_1>g_2$ always, thus $h_1$ has no zero either.

%% Then if $h(\rho)>0$ for all $\rho\in\mathbb{R}_+$ that means that $g$
%% never goes to $-\infty$.

\subsubsection*{Application 2}

We may apply the above to the case when Eq.~\eqref{eq:gen-2o} is taken
to be Eq.~\eqref{SmallPrEquationsb}.  For $P\to0$ the intial condition
at $\rho=0$ becomes $h_M'=0$.  The case $z=0$ can be explicitly solved
and we get $h_M(\rho) = \rho^{-d/2} J_{d/\xi}(2\sqrt{d/\xi+1} \,
\rho^{\xi/2})$.  What needs to be seen is that $h_M$ does not have
zeros between 0 and 1, equivalent to $J_{d/\xi}$ not having zeros
between 0 and $2\sqrt{d/\xi+1}$, which follows from the fact that
$j_{\nu,1} > 2\sqrt{\nu+1}$ for $\nu \geq 0$ (where $j_{\nu,1}$ is the
first positive zero of the Bessel function of index $\nu$), which may
be easily verified by a plot or by more serious analysis.  We remark
that $h_M'$ does have zeros, as an effect of the term $\xi(d-1)(d+\xi)
\rho^{\xi-2} h_M$ in Eq.~\eqref{SmallPrEquationsb}, and thus the fact
that $\Lambda_L$ is indeed monotonously increasing is surprisingly non
trivial {\em because} of this term.

All this allows us to conclude that
%% $h(\rho)$ has no zero for $\rho<1$ [seen only from plot?], so
$\Lambda_L = h_M'(1)/h_M(1)$ grows with $z$ (equivalently with $w$).

\subsubsection*{Application 3}

Along the same lines one can prove for our case the standard lore of
Sturm-Liouville theory that if the zero mode ($z=0$ solution) has no
zeros, then there is no eigenfunction with $z>0$.

The idea is that while the zero mode decays near infinity as a power
law, any eigenfunction $h_1$ for $z>0$ has to decay exponentially, so
it will be below the zero mode.  On the other hand, from
Eq.~\eqref{eq:gen-2o} on deduces that $h''(0)$ grows with $z$, so that
$h_1$ has to be larger than the zero mode near $\rho=0$.  This would
imply that the two have to cross in the sense that $h_2$ comes from
above and goes below the zero mode, but at the crossing point $g_1$
would be less than that of the zero mode, which contradicts the above
said.

%% Take $z>0$, then large $\rho$ asymptotics of the convergent solution
%% is such that it has to go below the $z=0$ convergent solution: in our
%% case more or less like $\exp(-\sqrt{z}\rho^{1-\xi/2})$, compared with
%% a power law for $z=0$.

\subsection{Consequences for modified Bessel function}
\label{sec:Sturm-Liouville:sub:Bessel}

We wish to prove here that, for pure imaginary $\lambda$, the
slope of $\rho K'_\lambda(\rho)/K_\lambda(\rho)$ is bounded from above
by $-1$ for all $\rho > 0$.

Using notation from the previous subsections, introduce $f(\rho) =
\rho g(\rho)$.  Then Eq.~\eqref{eq:g-prime} translates to $f' =
[(z-c)\rho+(a/\rho -b)f-(a/\rho)f^2]/a$.  Applied to the particular
case of the modified Bessel equation with parameter $\lambda$
$$
  \rho^2 h'' + \rho h' - \rho^2h
=
  \lambda^2 h
\,,
$$
i.e.\ when $a(\rho) = \rho^2$, $b(\rho) = \rho$, $c(\rho)=-\rho^2$ and
$z=\lambda^2$, we obtain
\begin{equation}
\label{eq:Bessel-f-prime}
  f'
=
  (\rho^2+\lambda^2-f^2)/\rho
\,.
\end{equation}

Solving Eq.~\eqref{eq:Bessel-f-prime} for $f'=-1$ gives $f^2 =
s(\rho)^2$ where we define $s(\rho) =
-[(\rho+1/2)^2+\lambda^2-1/4]^{1/2}$.  Moreover when $f^2 > s^2$ then
$f'<-1$.

We now take $f=\rho K'_\lambda(\rho)/K_\lambda(\rho)$ in the case when
$\lambda$ is pure imaginary.  Then, for large $\rho$, asymptotically
$f(\rho) - s(\rho) \sim -(1-4\lambda^2)/(16\rho^2) < 0$, the last
inequality being guaranteed by the fact that we consider the case when
$\lambda$ is pure imaginary and hence $\lambda^2 \leq 0$.  This means
that for large $\rho$ asymptotically $f < s$.

Using the fact that $f$ is continuous, if $f$ were to become larger
than $s$ for some finite $\rho$, necessarily it would pass through
$f=s$, but at that point we would have $f'=-1 > s'$ (the inequality
holding for $\lambda$ pure imaginary), which is a contradiction to the
fact that for larger $\rho$ we should have $f<s$.

This proves that $f<s \leq 0$ when $s$ is real, and thus $f^2>s^2$ for
all $\rho \geq 0$, whence $f'<-1$ for $\rho \geq 0$.

\subsection{Real spectrum}

Though we do not consider $\mathcal M$ to be self-adjoint, its
spectrum is always real, for the following reason.

Since $\mathcal M$ is a second order differential operator we may
conjugate it by a multiplication operator (by a ``function'' which is
known in the thory of diffusion processes as the speed measure) to get
a symmetric operator $\tilde{\mathcal M}$, and taking into account the
boundary conditions we have (we see that for any $z \in
\CC\setminus\RR_-$ the solution of $\tilde{\mathcal M} h = z h$ which
verifies the boundary conditions is a twice differentiable function
with zero derivative at $\rho=0$ and exponentially decaying as
$\rho\to\infty$, so $h$ is also in the domain of $\tilde{\mathcal
M}^\dag$), we can use the same trick as for self-adjoint operators:
suppose $\tilde{\mathcal M} h = z h$ and write $\int \bar h
\tilde{\mathcal M} h = z \int \bar h h$, now take the complex
conjugate of both sides, and since $\tilde{\mathcal M}$ is real and
symmetric, we have $\int \bar h \tilde{\mathcal M} h = \bar z \int
\bar h h$, showing that $z = \bar z$, i.e.\ that $z$ is real.

\section{Exact results for $P=0$}

Here we want to study more rigorously the case of $P=0$.  In this case
we can find exactly the zero mode of Eq.~\eqref{eka} and show that for
$\lambda$ real (recall its definition from Eq.~\eqref{lambda}) it has
no nodes.  On the other hand for $\lambda$ pure imaginary it has an
infinity of nodes.

Recalling Eq.~\eqref{eq:P0-match-ML}, first we have to solve the zero
mode equation $[\xi(d-1)(d+\xi)\rho^{\xi-2} + (d^2-1+2\xi)\rho^{\xi-1}
\partial_\rho + (d-1)\rho^\xi \partial_\rho^2] + [(d^2-1)\rho^{-1}
\partial_\rho + (d-1) \partial_\rho^2]$.  At 0 Prandtl number the
boundary condition is to have finite limit at $\rho=0$.  The
appropriate solution is
$$
  (\rho^\xi+1)^{(d-3)/(d-1)}
  {}_2F_1 \left(
    \begin{array}{c}
      \frac{2(d-2)\xi+d(d-1)+(2-\xi)(d-1)\lambda}{2\xi(d-1)},
      \frac{2(d-2)\xi+d(d-1)-(2-\xi)(d-1)\lambda}{2\xi(d-1)}\\[2ex]
      \frac{d+\xi}{\xi}
    \end{array};
    -\rho^\xi
  \right)
$$
where ${}_2F_1$ is the hypergeometric function, which we shall simply
denote as ${}_2F_1(a,b;c;x)$, and $a,b,c,x$ are defined accordingly to
the above displayed formula.

Let us start with the case of $\lambda$ real.  Without loss of
generality, we may suppose $\lambda>0$ (or otherwise exchange $a$ and
$b$, since the hypergeometric function is symmetric in those
arguments).  Notice that $2(d-2)\xi+d(d-1) > 0$ and
$[2(d-2)\xi+d(d-1)]^2-[(2-\xi)(d-1)\lambda]^2 = 2d^2(d-1)^2 -
8\xi^2(d-2) \geq 2[d^2(d-1)^2 - 16(d-2)] > 0$, implying $b>0$.

%% Hypergeometric identity ${}_2F_1(a,b;c;x) = (1-x)^{-a}
%% {}_2F_1(a,c-b;c,x/(x-1))$\\
Notice also $2(d-1)(d+\xi)-[2(d-2)\xi+d(d-1)] = d(d-1)+2\xi > 0$
implying $c-b > 0$.

%% Thus both for $-1<x<0$ and $x<-1$ the coefficients $a,b,c$ are
%% positive, implying positivity of ${}_2F_1$.

Now write the following integral representation of the hypergeometric
function:
$$
  {}_2F_1(a,b;c;x)
=
  \frac{\Gamma(c)}{\Gamma(b)\Gamma(c-b)}
  \int_0^1 \frac{t^{b-1}(1-t)^{c-b-1}}{(1-tx)^a} \,dt
\qquad
  (c>b>0)
$$
whence ${}_2F_1(a,b;c;x) > 0$ for any $x<1$ and $c>b>0$.

Since we have shown above $c>b>0$ and since our $x<0$, this proves
that the zero mode has no zeros, and hence there cannot be a dynamo
effect.

%% Thus the zero mode is positive, meaning that there are no growing
%% states, i.e.\ no dynamo.

For $\lambda$ pure imaginary it is possible to make a large $\rho$
development using the so called linear transformation formula
${}_2F_1(a,b;c;-x) = \frac{\Gamma(c)\Gamma(b-a)}{\Gamma(b)\Gamma(c-a)}
x^{-a} {}_2F_1(a,1-c+a;1-b+a;-1/x) +
\frac{\Gamma(c)\Gamma(a-b)}{\Gamma(a)\Gamma(c-b)} x^{-b}
{}_2F_1(b,1-c+b;1-a+b;-1/x)$.  Since $a$ and $b$ are complex
conjugates in the case of pure imaginary $\lambda$, the large $x$
asymptotics can be written as ${}_2F_1(a,b;c;-x) \sim \Re
(\frac{\Gamma(c)\Gamma(b-a)}{\Gamma(b)\Gamma(c-a)} x^{-a})$, which has
an infinity of zeros since $a$ has an imaginary part.

\end{document}